\documentclass[journal,10pt,final]{IEEEtran}

\newcommand{\figwidth}{3.5in}

\usepackage{paper_header}

\usepackage{float}
\usepackage[displaymath, tightpage]{preview}
\usepackage{natbib}
\renewcommand{\cite}{\citep}		

\begin{document}

\title{%
Patterns of Ship-borne Species Spread: A Clustering Approach for Risk Assessment and Management of Non-indigenous Species Spread
\author{%
Jian Xu$^*$, Thanuka Wickramarathne$^{*\dagger}$, Erin Grey, Karsten Steinhaeuser, Reuben Keller, John Drake, \\Nitesh Chawla and David Lodge}
\thanks{%
${}^*$JX and TW have contributed equally; ${}^{\dagger}$TW is the corresponding author.}
\thanks{%
Authors JX, TW and NC are with Dept. of Computer Science and Engineering; EG and DL are with Dept. of Biology of University of Notre Dame (ND), Notre Dame, IN, USA. Authors TW, EG, NC and DL are also with ND Environmental Change Initiative (ECI);  Authors JX, TW, NC and DL are with ND interdisciplinary Center of Network Science and Applications (iCeNSA). Authors KS, RK and JD were previously with ND before joining Dept. of Computer Science and Engineering, University of Minnesota, the Institute of Environmental Sustainability, Loyola University Chicago and Odum School of Ecology, University of Georgia, respectively.}
\thanks{%
    This work is based on research supported by the ND Office of Research via funding under ND ECI for TW, EG, NC and DL.}}

\maketitle

\begin{abstract}
The spread of non-indigenous species (NIS) through the global shipping network (GSN) has enormous ecological and economic cost throughout the world. Previous attempts at quantifying NIS invasions have mostly taken ``bottom-up'' approaches that eventually require the use of multiple simplifying assumptions due to insufficiency and/or uncertainty of available data. By modeling implicit species exchanges via a graph abstraction that we refer to as the \emph{Species Flow Network (SFN),} a different approach that exploits the power of \emph{network science} methods in extracting knowledge from largely incomplete data is presented. Here, coarse-grained species flow dynamics are studied via a \emph{graph clustering approach} that decomposes the SFN to \emph{clusters of ports} and \emph{inter-cluster} connections. With this decomposition of ports in place, NIS flow among clusters can be very efficiently reduced by enforcing NIS management on a few chosen inter-cluster connections. Furthermore, efficient NIS management strategy for species exchanges within a cluster (often difficult due higher rate of travel and pathways) are then derived in conjunction with ecological and environmental aspects that govern the species establishment. The benefits of the presented  approach include robustness to data uncertainties, implicit incorporation of ``stepping-stone'' spread of invasive species, and decoupling of species spread and establishment risk estimation. Our analysis of a multi-year (1997--2006) GSN dataset using the presented approach shows the existence of a few large clusters of ports with higher intra-cluster species flow that are fairly stable over time. Furthermore, detailed investigations were carried out on vessel types, ports, and inter-cluster connections. Finally, our observations are discussed in the context of known NIS invasions and future research directions are also presented.  
\end{abstract}


\section{Introduction}

Commercial shipping provides enormous economic benefits worldwide and is responsible for approximately 90\% of global trade. However, shipping also imparts large economic and environmental costs by spreading invasive species, or those non-indigenous species (NIS) that damage ecological systems. Shipping can translocate NIS to new areas either through ballast water or hull-fouling, and is responsible for 69\% of known aquatic NIS \cite{molnar_assessing_2008}. Although only a small portion of transported NIS establish and become invasive, their environmental and economic damages are often large and grow over time~\cite{halpern_global_2008,keller_bioeconomics_2009}. For instance, we recently estimated that ship-borne aquatic invasive species cost the Great Lakes regional economy \$100--800 million annually~\cite{rothlisberger_ship-borne_2012}. This high cost of ship-born invasions has motivated several efforts to better understand NIS spread and invasion risk through the global shipping network (GSN)~\cite{drake_global_2004,kaluza_complex_2010,Kel11,kolzsch_indications_2011,See13}. These studies used ship traffic data to create a \emph{network}, where \emph{nodes} (i.e., ports) are connected by \emph{edges} that represent the intensity of shipping traffic. Such networks have been shown to have \emph{small-world}~\cite{Wat98} characteristics, wherein each port is linked to any other port by a small number of ``hops''~\cite{kaluza_complex_2010, Kel11, kolzsch_indications_2011}, and to be very robust with many redundant links~\cite{kaluza_complex_2010}.

While such initial network analyses are enlightening, they are ultimately inadequate because ship traffic cannot sufficiently capture NIS invasion risk. Rather, invasion risk is likely affected by a complex interplay of ship traffic, ballast uptake/discharge dynamics, survival during transport, propagule pressure, environmental variables, biotic interactions and several other variables that are not yet well characterized~\cite{Won13}. Incorporating these complexities is a challenging task, since majority of the above relationships and their parameterizations are poorly known. The few studies that have attempted to calculate more realistic measures of invasion risk have relied on probabilistic models that make several simplifying assumptions. For example,~\cite{Kel11} combined ship traffic and environmental similarity to estimate relative invasion risk, assuming that simple Euclidean distance between ports' mean annual temperature and salinity was proportional to risk. This linear relationship between risk and changes in temperature and salinity is not likely for most species, particularly invasive species who tend to exhibit broad environmental tolerances. Most recently,~\cite{See13} calculated between-port invasion risk as the product of three probabilities---the probability a species was non-native (based on geographic distance), the probability a species survived transport (based on trip duration), and the probability a species establishes (based on Euclidean environmental similarity). The benefits of these probabilistic approaches are that they provide quantitative estimates. Their drawbacks include unjustifiable simplifying assumptions (i.e., establishment proportional to Euclidean distance, linear propagules pressure~invasion risk relationships), high uncertainty, and inability to incorporate ``stepping-stone'' invasion probabilities. 

The graph analysis methods popularized by network science are excellent tools for our goals, as they provide some of the most elegant tools for descriptive analysis of complex, relational data, 
and they are able to reveal large-scale patterns from a higher level, which is not easily affected by small uncertainties in data.

Specifically, we 1) create a network that represents the general species flow tendency among ports, 2) identify \emph{clusters}, or groups of ports, in which intense species flow tightly connect the ports in the same cluster, while connections between different clusters are loose, 3) identify ports and ship types that serve as important ``inter-cluster connectors'', 4) develop flexible methods to qualitatively assess invasion risk within a cluster based on realistic biogeographic and environmental relationships, and 5) highlight the management implications of our results. We focus here on the spread of species via ballast water, but the method could be easily applied to hull-fouling spread with a few adjustments.


%
This paper is organized as follows: 
	Section~\ref{sec:mat_methods} presents the materials and methods describing the formulation of species flow networks using limited available data, graph clustering approach for understanding the large-scale dynamics of GSN, and an intuitive method that extends graph clustering notions for detailed risk analysis using ecoregion and environmental conditions; 
	Section~\ref{sec:results_discuss} presents the results and provides a detailed discussion;
    and finally, 
	Section~\ref{sec:conc} contains the concluding remarks.

\section{Materials and Methods}
\label{sec:mat_methods}

Our main goal is to understand the large-scale (or coarse-grained) patterns of GSN in order to obtain better insight towards ship-borne NIS invasions. The presented approach is developed in order to exploit the power of network analysis methods in extracting knowledge from largely incomplete data with minimal simplifications and assumptions. We proceed as follows:
	\tb{(i)} a network that represents the general species flow tendency among ports is built; 
then, utilizing a graph clustering method~\cite{Ros08} that operates on the basis of flow-dynamics,  
	\tb{(ii)} a \emph{map}~\cite{Gui05,Tuf06} of the species flow network, i.e., a cogent representation that extracts the main structure of flow while retaining information about relationships among modules (of main structure), is built;  
finally, using this map that summarizes the species flow dynamics in terms of \emph{clusters} (or groups) of ports and highlights \emph{inter-cluster} (i.e., between clusters) and \emph{intra-cluster} (i.e., within cluster) relationships,
	\tb{(iii)} the impact of GSN dynamics on NIS invasions is studied in conjunction with ecological and environmental aspects that govern the species establishment. 
Let us now illustrate this innovative approach in detail.

\subsection{Datasets and Other Information Sources}
\label{ssec:datasets}

\subsubsection{LMIU database}
Global and domestic vessel movements for four (4) periods of 1997--1998, 1999--2000, 2002--2003 and 2005--2006, totaling $6,889,748$ individual voyages corresponding to a total of $50,487$ vessels of various types that move between a total of $5,545$ ports and regions, are acquired from Lloyd's Maritime Intelligence Unit (LMIU). For each period, the LMIU database contains travel information for vessels such as {\tt portID}, {\tt sail\_date} and {\tt arrival\_date}, along with vessel metadata, such as {\tt vessel\_type} and {\tt DWT} (i.e., dead weight tonnage), etc. 

\subsubsection{NBIC database}
Since vessel movement data (including LMIU) does not provide explicit ballast water exchange amounts (or even whether a vessel dis/charged ballast water), these quantities must be estimated based on some auxiliary data that can sufficiently relate ballast discharge to vessel information given in the LMIU database. Therefore, we utilize the approach suggested in~\cite{See13}, where ballast water discharge amounts are calculated using a linear regression model per {\tt vessel\_type} basis. For this, the National Ballast Water Clearinghouse (NBIC) database, which contains the {\tt date} and {\tt discharge\_volume} of all ships visiting U.S. ports from Jan. 2004 to present, is used (see Section~\ref{para:bd_estimate} for details).

\subsubsection{Ecoregion and environmental data}

Ecoregions are defined by species composition and shared evolutionary history~\cite{Spa07}, and thereby capable of providing a more realistic outline of native and invasive ranges. Therefore, we define non-indigenous status based on \emph{ecoregion concept} in comparison to, for example, geographic distance as used in~\cite{See13}. Here, ecoregion delineations given by Marine Ecoregions of the World~\cite{Spa07} and the Freshwater Ecoregions of the World~\cite{Abe08} are used. Then, annual averages of port temperature and salinity are given in the Global Ports Database (GPD)~\cite{Kel11} are used for assessment of NIS establishment risk that is based on environmental similarity; the missing values in GPD are supplemented by estimates from the World Ocean Atlas 2009~\cite{WOA09_sal,WOA09_temp} when necessary.

\subsection{Network Modeling for Species Flow Analysis}
\label{ssec:nis_network}

At the heart of a network analysis lies a graph abstraction of the (often complex) system that is under investigation. This graph must be capable of adequately capturing the system behavior via sets of \emph{nodes} and \emph{edges} that model flow/connectivity characteristics. Previous work [@cite] on analysis of GSN impact on NIS invasions, have employed \emph{undirected weighted graphs,} where nodes are given by the ports (visited by GSN) and edge (and their strength or weight) are derived from traffic intensity between ports. While such modeling is perhaps adequate for network analysis, the task at hand, viz., an analysis based on flow dynamics, a \emph{directed} network that can adequately represent the directional and asymmetric flow between nodes is mandatory. Therefore, a directed weighted graph that we refer to as the \emph{Species Flow Network (SFN)} is derived to better represent species flow characteristics among ports. Here, species flow is derived based \emph{only} on ballast water exchange, and contribution from hull-fouling is not considered. Therefore, the resulting flow dynamics represent the species flow with respect to ballast exchange only. Investigation of bio-fouling is relegated to a future publication (see Section~\ref{sec:conc}).

\begin{figure}[ht!]
\begin{center}
\includegraphics[width=\figwidth]{%
	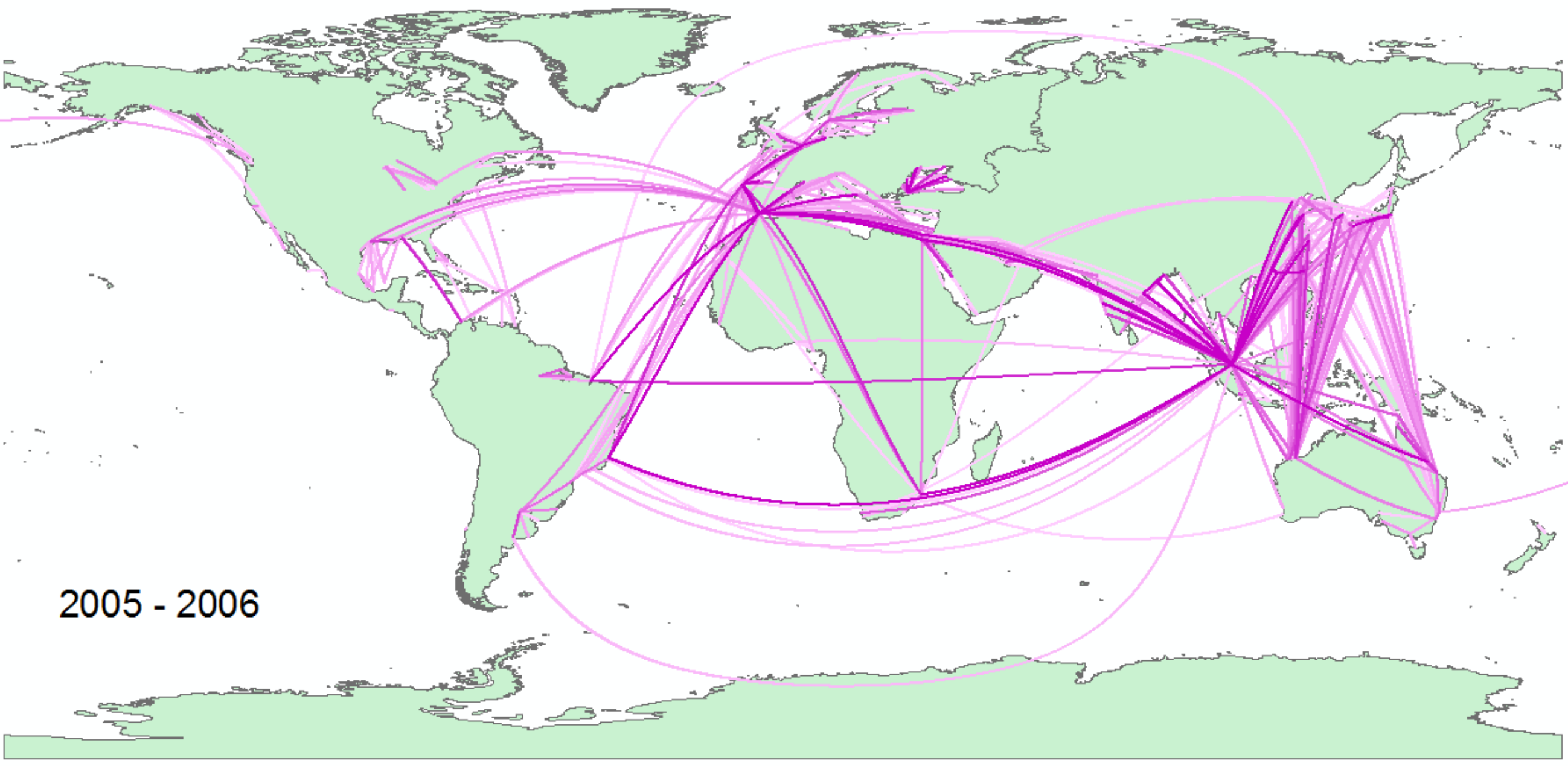}
\vspace*{-0.15in}
\end{center}
\vspace*{-0.2in}
\caption{\scriptsize\tb{Species flow between ports corresponding to vessel movements given in the LMIU 2005--2006 dataset.} The edges represent the aggregated species flow between ports, where the color intensity is proportional to the magnitude of flow. Approximately, $2300$ paths with the highest species flow are shown.}
\label{fig:nis_flow}
\vspace*{-0.1in}
\end{figure}

\subsubsection{Species Flow Network (SFN)}
Consider a \emph{directed graph} $\mc{G}\equiv(\mc{N},\mc{E})$, where $\mc{N}\equiv\{n_1,\ldots,n_n\}$ and $\mc{E}\subset \mc{N}\times\mc{N}$ denote the set of nodes and edges of $\mc{G}$, respectively. Let the nodes in $\mc{N}$ correspond to ports visited by the GSN and the \emph{weight} of the directed edge $e_{ij}\in \mc{E}$ given by $w_{ij}\in(0,1]$ represents the \emph{total probability of species introduction} corresponding to all vessels traveling from port $n_i$ to $n_j$ (without intermediate stopovers), for all $n_i,n_j\in\mc{N}$. 

\paragraph{Estimation of species flow}

Species flow between two ports is estimated as in~\cite{See13}. To summarize, consider a vessel $v$ traveling from port $n_i$ to $n_j$ (without intermediate stopovers) in $\Delta t^{(v)}_{ij}$ time, during which the species in ballast water die at a \emph{mortality rate} of $\mu$ (which is set to a constant average of $0.02/day$ for all routes $r$ and vessel types in experiments). In addition, let $D^{(v)}_{ij}$, $\rho^{(v)}_{ij}\in[0,1]$ and $\lambda$ denote the amount of ballast water discharged by vessel $v$ at $n_j$, the efficacy of ballast water management for $v$ for the route $n_i\to n_j$, and the characteristic constant of discharge, respectively. Then, the probability of vessel $v$ introducing species from $n_i$ to $n_j$ (without intermediate stopovers) is given by: 
\begin{equation}
\label{eq:pv_ij}
p^{(v)}_{ij} = \rho^{(v)}_{ij}\,(1- e^{-\lambda D_{ij}^{(v)}})\,e^{-\mu \Delta t_{ij}^{(v)}};
\end{equation}
then, the total probability of species introduction for all vessels traveling from $n_i$ to $n_j$ is given by:
\begin{equation}
\label{eq:w_ij}
w_{ij} = 1 - \prod_{\substack{r\in DB\\ r=v:n_i\rightarrow n_j}} (1 - p^{(v)}_{ij}),
\end{equation}
where the product is taken over all routes $r$ in database $DB$ s.t. a vessel $v$ travels from port $n_i$ to $n_j$.

\paragraph{Estimation of ballast discharge}
\label{para:bd_estimate}
Information on ballast dis/charge are largely incomplete to a degree, where estimation of exact quantities exchanged for each and every ship route $r$ is almost impossible due to numerous reasons: 
	\tb{(i)} ballast dis/charges in ports are not recorded globally, and are known to vary significantly by port and ship type; 
	\tb{(ii)} vessels may have intermediate stopovers, thus exchanging and mixing ballast water with existing water in ballast tanks; and 
	\tb{(iii)} data are largely unavailable for offshore discharges.
Therefore, in order to mitigate the above difficulties, ballast discharge is estimated based on linear regression models on {\tt DWT} per {\tt vessel\_type} as in~\cite{See13}. Specifically, linear regression models on {\tt DWT} for vessels of type {\tt Bulk Dry}, {\tt General Cargo}, {\tt Ro-Ro Cargo}, {\tt Chemical}, {\tt Liquified Gas Tankers}, {\tt Oil Tankers}, {\tt Passenger Vessels}, {\tt Refrigerated Cargo}, {\tt Container Ships} and {\tt Unknown/Other}) are derived using only the non-zero discharge events recorded in NBIC database. 

Furthermore, the relationship of ballast discharge amount to the likelihood of species introduction is not well defined. Therefore, for estimation of \eqref{eq:pv_ij}, $\lambda$ is chosen s.t. $p^{(v)}_{ij}=0.80$ for a ballast discharge of $500,000\,m^3$, when $\rho^{(v)}_{ij}=1$ and $\Delta t_{ij}^{(v)}=0$, i.e., a discharge volume of $500,000\,m^3$ has a probability of $0.8$ of introducing species if the vessel travels with zero mortalities and has no ballast management strategies in place.


\paragraph{Network characteristics}
Table~\ref{tb:net_char} summarizes the characteristics for SFNs generated for the four (4) LMIU datasets.

\begin{table}[ht!]
\vspace*{-0.1in}
\scriptsize
\renewcommand{\arraystretch}{1.1}\addtolength{\tabcolsep}{0pt}
\caption{\tb{Characteristics of Species Flow Networks}}
\vspace*{-0.1in}
\begin{tabular}{l|rrrr}
\hline
							&\tb{1997--1998}		&\tb{1999--2000}		&\tb{2002--2003}		&\tb{2005--2006}\\
\hline
Number of nodes 			&3971		&4045		&4264		&4250\\
Number of edges			&150479	&150150	&143560	&145199\\
Average path length		&2.987		&2.998		&3.018		&3.041\\
Average in/out- degree		&37.9		&37.1		&33.7		&34.2\\
Diameter					&8			&7			&7			&9\\
Density						&0.010		&0.009		&0.008		&0.008\\
\hline
\end{tabular}
\label{tb:net_char}
\end{table}

The \emph{path length} of a network identifies the number of stops required to reach a given port from another. An \emph{average path length} of three (3) is observed in all four SFNs, indicating that they fall under the category of \emph{small-world networks}~\cite{Wat98}. This is perhaps mainly due to the presence of \emph{hubs} (i.e., ports that are connected to many other ports) in GSN (e.g., Singapore). The \emph{in-/out-degree} of a node is defined as the number of other nodes connected to/by it. Therefore, \emph{average degree} in SFN describe the average number of pathways of species introduction. Furthermore, as characterized by the \emph{power law degree distribution}~\cite{Cla09}, SFN also falls under the category of \emph{scale-free networks}~\cite{Bar00}. Moreover, the scale-free nature of SFN indicates the possibility of existence of dense subgraphs connected by hubs, which we exploit as follows in our analysis.

\subsection{Clustering Analysis of Species Flow Network}
\label{ssec:clustering}
\begin{figure}[ht!]
\centering
\includegraphics[width=3.5in]{%
	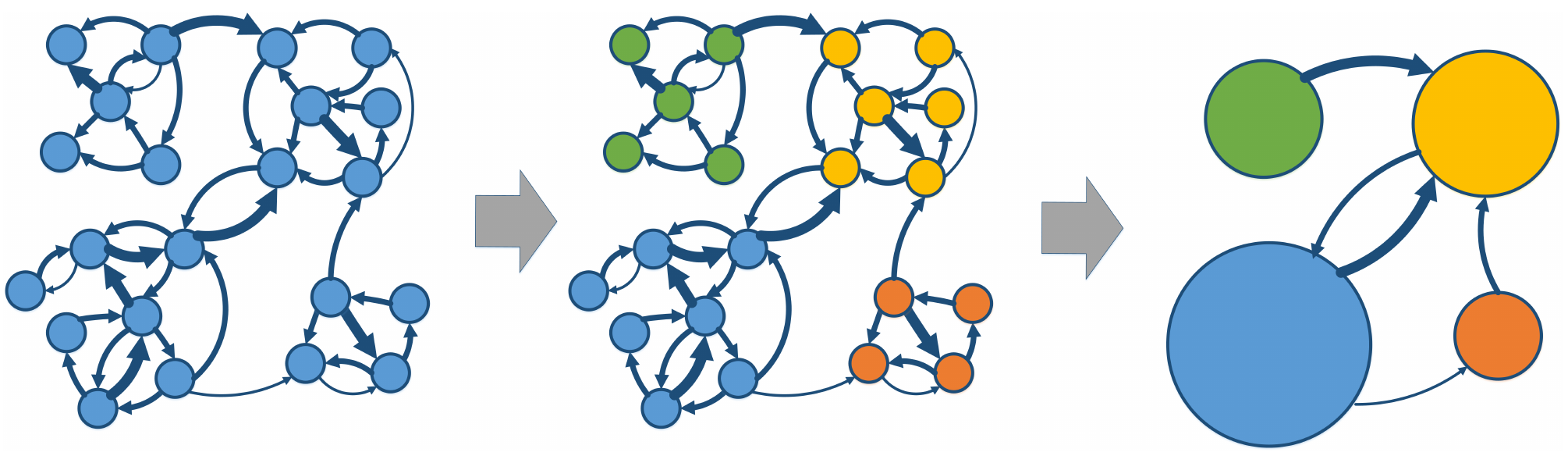}
\vspace*{-0.25in}
\caption{\scriptsize\tb{Illustration of clustering analysis of SFN.} SFN is first decomposed into dense sub-graphs referred to as \emph{clusters} based on flow characteristics; these clusters of ports are generated s.t. the species flow between two ports that belong to the same cluster is higher than the flow between ports that belong to two different clusters; then, the clusters and inter-cluster connections provide a higher level description of the flow dynamics of the SFN that we utilize later on for further analysis of species exchange among ports and ballast management strategies.}
\label{fig:network_to_clusters}
\vspace*{-0.1in}
\end{figure}

Complex networks are efficient abstractions for highly complex systems that consists of numerous, often complex underlying patterns and relationships. However, these abstractions still remain too complex to derive useful inferences. Therefore, a decomposition that represents such complex networks via \emph{modules} and their interactions~\cite{Gir02, Pal05, Par07} can be very useful in understanding the underlying patterns. As the network characteristics of the SFN suggests the existence of dense sub-graphs, we utilize a graph clustering approach in order to simplify the underlying flow dynamics of SFN (see Fig.~\ref{fig:network_to_clusters} for illustration of this concept).

The basic idea behind our analysis is to identify patterns of species flow, ports that are responsible for high species exchange and other auxiliary information (e.g., vessel types that are responsible for certain types of NIS invasions) in order to provide the knowledge that is required to devise and deploy NIS management strategies in a targeted and controlled manner. As shown in Fig.~\ref{fig:network_to_clusters}, a decomposition of SFN in terms of clusters of ports and inter-cluster pathways provide us with a higher level description of the GSN that can be utilized for this purpose. For instance, inter-cluster species flow pathways can be easily targeted to prevent NIS propagation between clusters, thus significantly reducing the number of pathways to be managed. 

For the task at hand, we are interested in understanding how the structure of SFN relates to species flow across the network. Therefore, among many alternatives, \emph{MapEquation}~\cite{Ros08}---a graph clustering method that attempts to decompose the network with respect to flow-dynamics (in comparison to optimization of \emph{modularity})---is used. The basic principle of operation behind MapEquation-based clustering stems from the notions of information theory, which states the fact that a data stream can be compressed by a \emph{code} that exploits regularities in the process that generates the stream~\cite{Sha63}. Therefore, a group of nodes among which information flows quickly and easily can be aggregated and described as a single well connected module; the links between modules capture the avenues of information flow between those modules. 

\vspace*{0.1in}
\noindent
\tb{Remark:}
MapEquation identifies clusters by optimizing the entropy corresponding to intra- and inter-cluster in a \emph{recursive} manner---the clusters identified cannot be further refined or partitioned. Therefore, SFN clusters in fact correspond to grouping of ports based on flow dynamics in comparison to a partition of ports into (specified) $n$ groups.

\subsection{Ecoregion and Environmental Considerations for NIS Invasion Risk Analysis}
\label{ssec:env}
Quantification of NIS invasion risk is a challenging problem because of the complex interactions between species and their abiotic and biotic environment~\cite{Won13}. Previous studies have assumed that the invasion risk is proportional to \emph{Euclidean} distance between annual averages of temperature and salinity~\cite{Kel11,See13,Flo13}. However, this assumption is likely not valid for most species, particularly for invasive species that often exhibit broad environmental tolerances~\cite{Goo99,Sim07}. Here we take a simple, yet intuitive approach that is based on biogeographic patterns and empirically observed temperature and salinity tolerances for ranking NIS invasion risks.

\subsubsection{NIS invasion risk}
\label{sssec:nis_invasion_risk}
between two ports is defined in two steps: 
	\tb{(i)} determine non-indigenous exchange status, and 
	\tb{(ii)} rank invasion risk based on environmental tolerance.

\begin{figure}[ht!]
\vspace*{-0.1in}
\begin{center}
\includegraphics[width=\figwidth]{%
  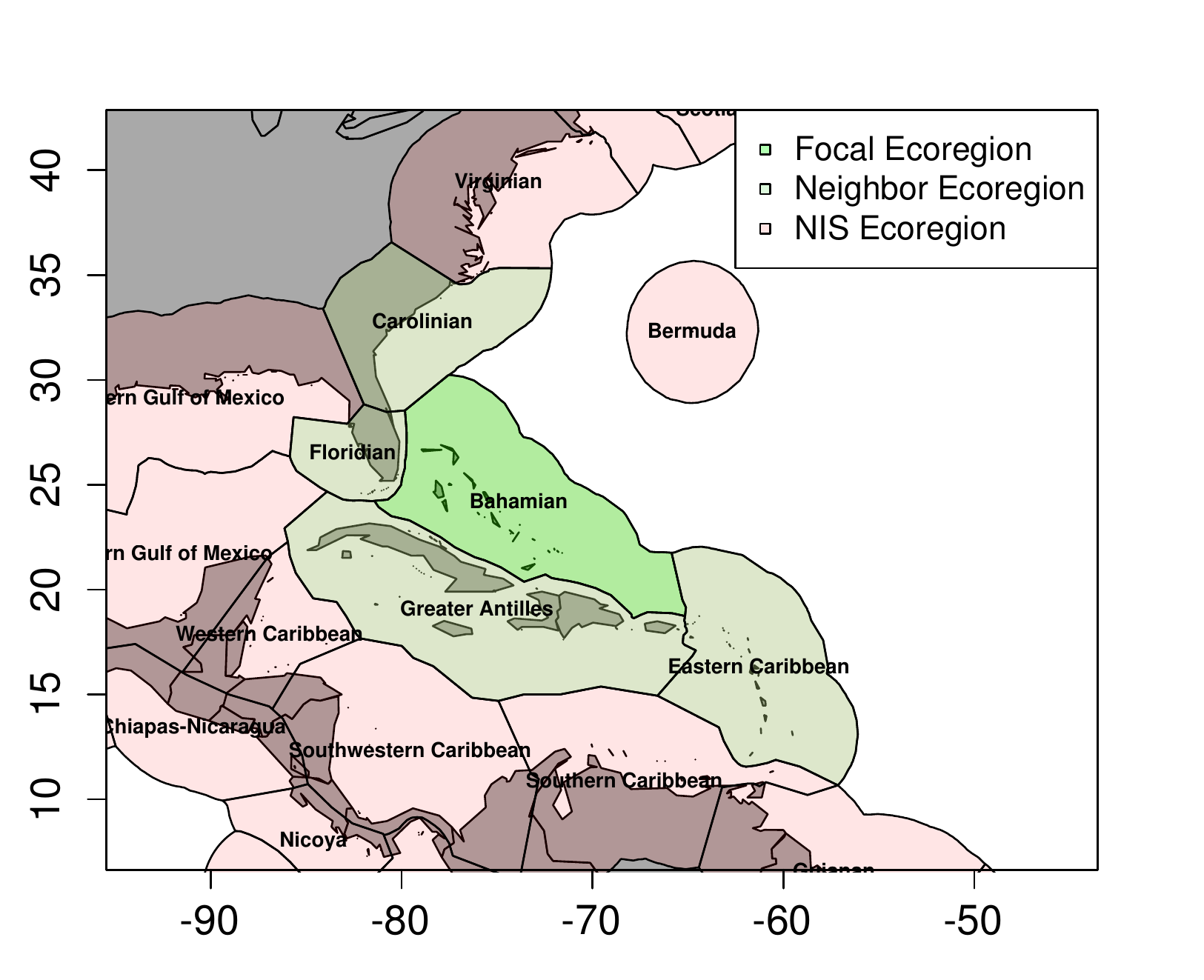}
\end{center}
\vspace*{-0.2in}
\caption{\scriptsize\tb{Illustration of non-indigenous exchange definition.} Ecoregions that are considered as non-indigenous exchanges with respect to the Bahamian ecoregion (Green) are shown. Contiguous neighbor ecoregions (light Green), such as the Floridian ecoregion are not considered as non-indigenous exchanges, while all non-contiguous ecoregions (light Red), such as Virginian ecoregion would be considered as non-indigenous exchanges. Ecoregion delineations and names from obtained from~\cite{Spa07}.}
\vspace*{-0.1in}
\label{fig:ecoregion}
\end{figure}

\paragraph{Non-indigenous exchange status}
is defined with respect to ecoregion concept. Given the fact that species can naturally disperse between contiguous ecoregions, we utilize a conservative NIS exchange status definition where only movements between non-contiguous ecoregions are considered as non-indigenous exchanges (see Fig.~\ref{fig:ecoregion} for example). 

\begin{figure}[ht!]
\begin{center}
\subfigure[Species tolerance groups]{
\includegraphics[width=3in]{%
	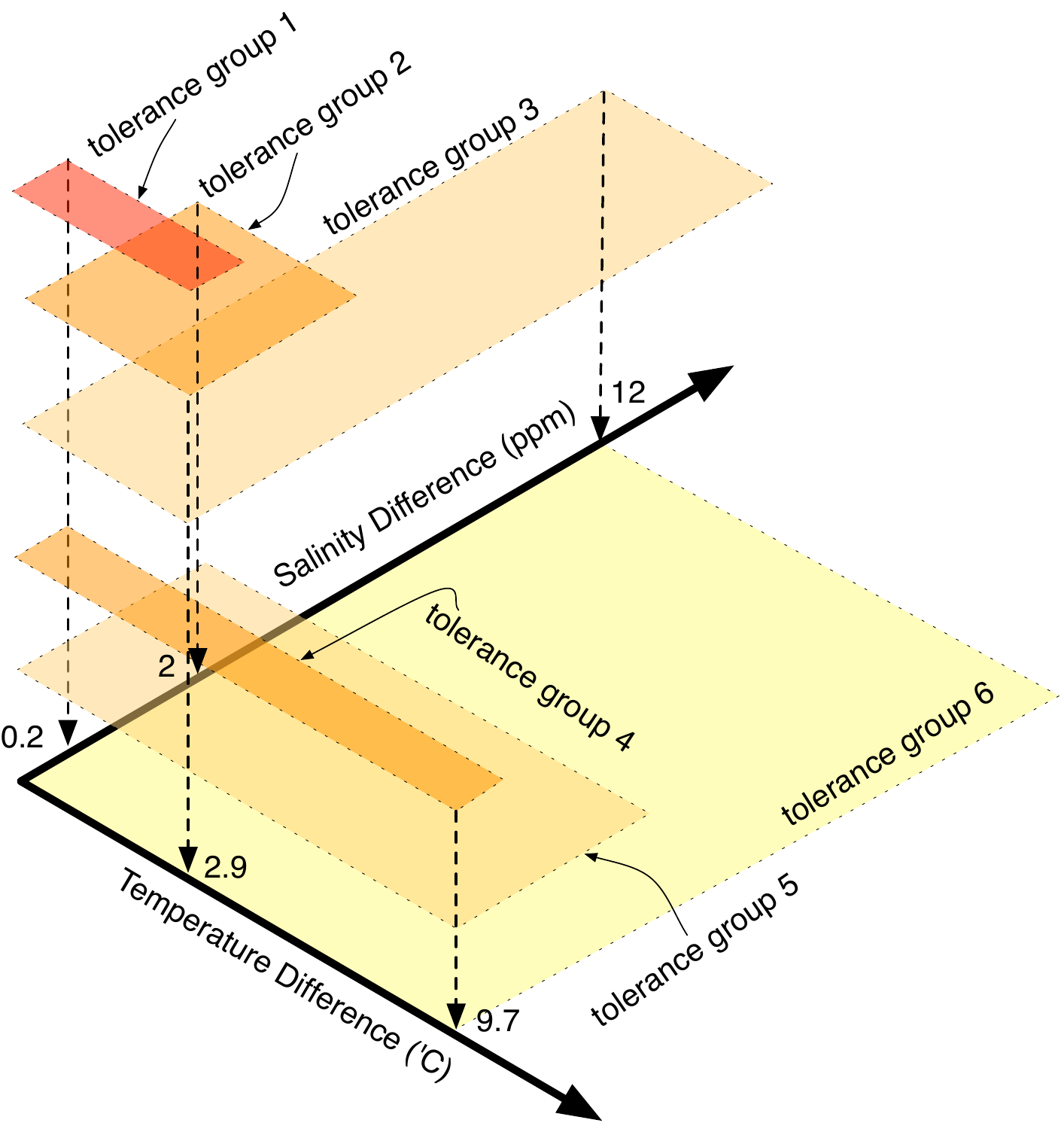}}
\subfigure[Risk level definition]{
\includegraphics[width=3in]{%
	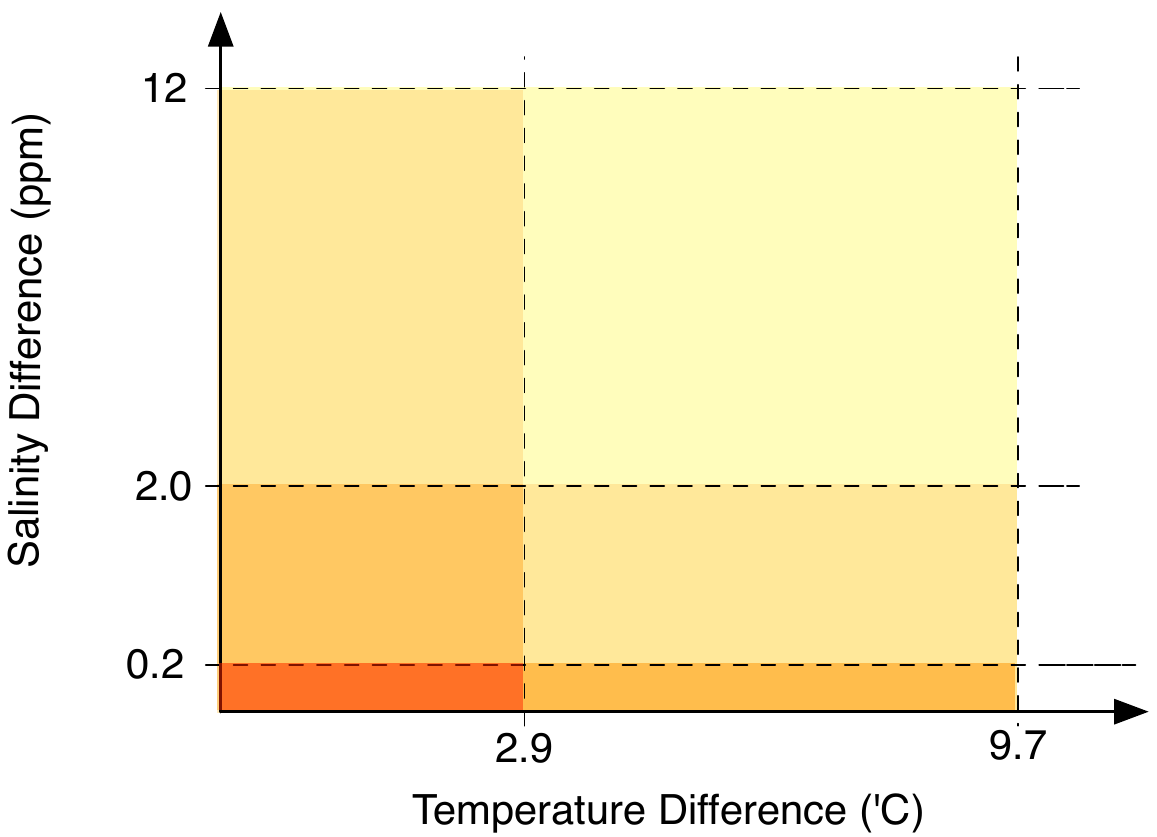}}
\end{center}
\vspace*{-0.1in}
\caption{\scriptsize\tb{Illustration of \emph{risk level} definition based on species tolerance groups and between-port environmental differences.} Sub-figure (a): number of species types at risk, identifies six (6) different species groups that may be at risk based on two (2) temperate tolerance levels (high =  can survive up to $9.7^{\circ}C$ and low = can survive up to $2.9^{\circ}C$ temperature difference) and three (3) salinity tolerance levels (zero = $0.2ppm$, low = $2.0ppm$ and high = $12.0ppm$ tolerance). Sub-figure (b): definition of risk level, defined based on number of species groups at risk as identified in (a); the colors are generated by overlapping the layers and later enhanced for clarity and ease of distinction. In this setting, risk level ranges from 0 to 6.}
\label{fig:risk_level}
\vspace*{-0.2in}
\end{figure}

\paragraph{Invasion risk between port environments}
is ranked by considering a species assemblage that contains ``generalist'' and ``specialist'' species. Specifically, six (6) different species tolerance groups based on two (2) temperature and three (3) salinity tolerance levels are considered (see Table~\ref{tab:tol_groups}). Here, temperature tolerance levels were set on empirically-estimated long term thermal tolerances for temperate marine invertebrate taxa~\cite{Ric12}; salinity tolerance levels were set to capture species types that are completely intolerant to salinity (i.e., freshwater species), those that are restricted to marine waters (i.e., low tolerance), and estuarine species that can survive in a wide range of salinities (i.e., high tolerance). Risk between any two port pairs is then quantified as an index created by overlapping the species tolerance groups as illustrated in Fig.~\ref{fig:risk_level}. 

\begin{table}[ht!]
\renewcommand{\arraystretch}{1.1}\addtolength{\tabcolsep}{6pt}
\caption{\tb{Species grouping based on environmental tolerance}}
\vspace*{-0.1in}
\centering
\begin{tabular}{l|cc}
\hline
Species Tolerance Group & \mcol{2}{c}{Tolerance Levels}\\
						 	& $\Delta T\,(^{\circ}C)$ 		& $\Delta S\,(ppt)$\\
\hline
Tolerance Group 1				& [0, 2.9]			& [0, 0.2]\\
Tolerance Group 2				& [0, 2.9]			& [0, 2.0]\\
Tolerance Group 3				& [0, 2.9]			& [0, 12]\\
Tolerance Group 4				& [0, 9.7]			& [0, 0.2]\\
Tolerance Group 5				& [0, 9.7]			& [0, 2.0]\\
Tolerance Group 6				& [0, 9.7]			& [0, 12]\\
\hline
\end{tabular}

\noindent{\scriptsize Tolerance groups are defined by temperature and salinity tolerance levels of species. For example, tolerance group 1 identifies all species that can survive a difference of $0-2.9^{\circ}C$ in temperature and $0-0.2\;ppt$ in salinity.}
\label{tab:tol_groups}
\end{table}

\subsubsection{Intra-cluster NIS invasion risk analysis}
For an exchange of species to become a NIS invasion, the introduced species must be
	\tb{(i)} a non-indigenous exchange, and 
	\tb{(ii)} able to survive and establish in their new environment. 
Assuming that inter-cluster species flow can be controlled by adequate ballast management on inter-cluster pathways, let us now focus on intra-cluster (i.e., ports within a cluster) NIS invasion risk in order to gain insight into plausibility of invasions in terms of environmental similarity, assuming that species exchange within a cluster is high enough to exert sufficient \emph{propagule pressure}~\cite{Won13} for species establishment.

\begin{figure}[ht!]
\begin{center}
\includegraphics[width=3.5in]{%
	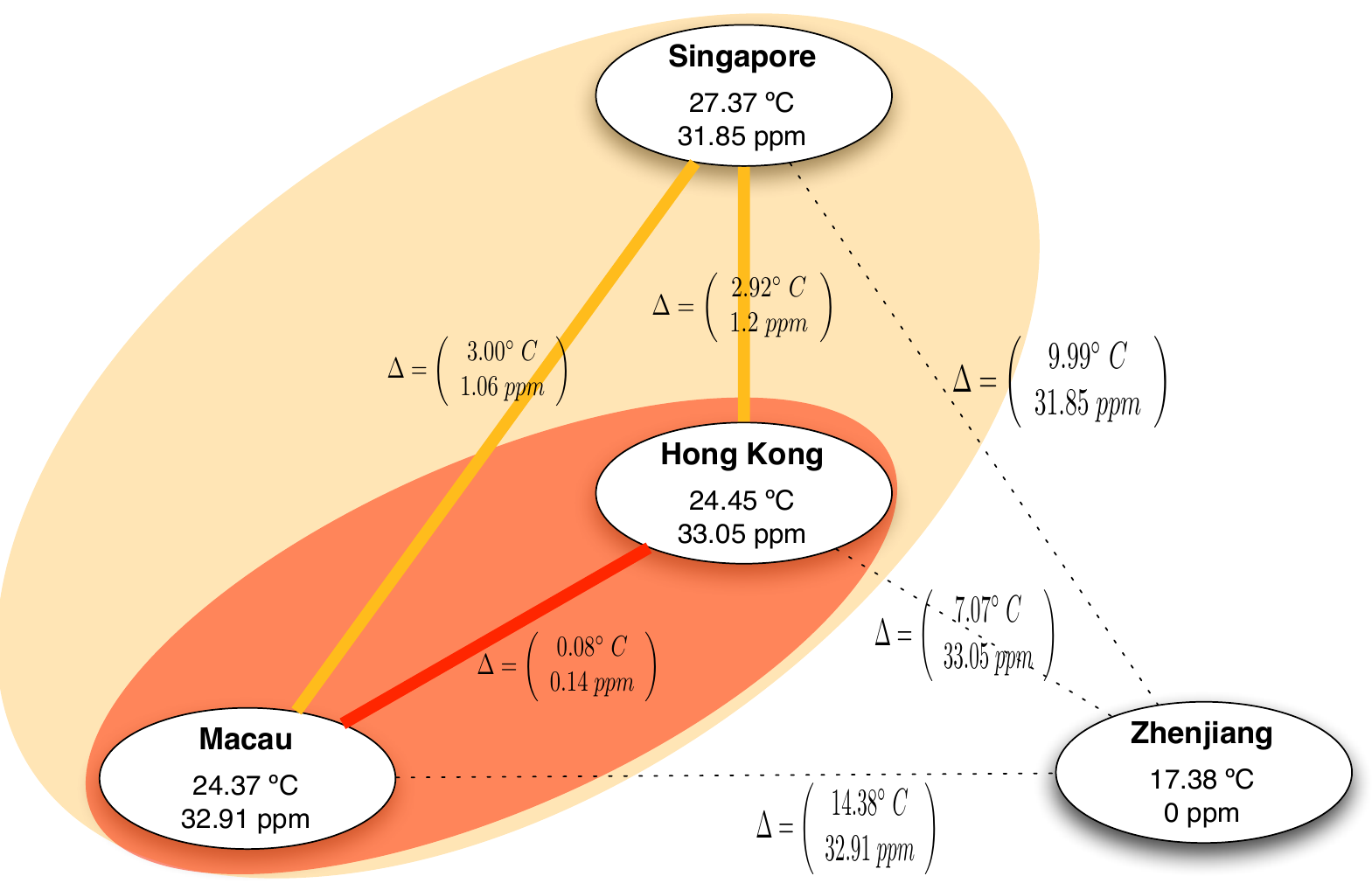}
\vspace*{-0.35in}
\end{center}
\caption{\tb{Illustrating the generation of NIS Invasion Risk Network (NIS-IRN).} NIS-IRN is an undirected graph where nodes and edges are given by the ports visited GSN and NIS invasion risk level, respectively. Shown here are four ports along with annual average temperature and salinity, and pair-wise salinity and temperature differences. For edges drawn in solid lines represent the risk level between ports, as defined in Fig.~\ref{fig:risk_level}; dotted-lines show zero (0) risk edges; colored-patches are used to show the overlap of species tolerance groups shared by a port-pair.}
\label{fig:env_network}
\vspace*{-0.1in}
\end{figure}

\paragraph{A network for port environmental similarity}
A graphical representation that we refer to as \emph{NIS Invasion Risk Network (NIS-IRN)} is built for every \emph{major} cluster in SFN to intuitively represent the NIS invasion risk between ports. NIS-IRN represents 
	\tb{(i)} whether species exchanges between two ports satisfy non-indigenous exchange status, and 
	\tb{(ii)} how easily they can establish in the new environment. 
For every port pair, the edge weight indicates the NIS invasion risk as identified above (see Fig.~\ref{fig:risk_level}); a zero (0) risk level correspond to species exchanges, where they will not survive due to mismatch of environmental conditions, and vice versa; edges are removed (hence indicating zero invasion risk) from port pairs that belong to the same or adjacent ecoregions, in order to represent \emph{only} non-indigenous species exchanges. See Fig.~\ref{fig:env_network} for an illustration. Note that NIS-IRN is both undirected and weighted, since environmental match is symmetric, and the risk level (based on number of tolerance groups at risk) can vary between port pairs, respectively. 

\paragraph{Clustering Analysis of NIS-IRN}
With the edges representing the NIS invasion risk between ports, clustering in this scenario can help detect groups of ports that have similar environmental conditions (while belonging to different and non-neighboring ecoregions). The basic idea here is to exploit the fact that the NIS invasion risk between groups of ports that are very dissimilar (e.g., fresh-water ports and marine ports) is lower than ports within the same group (with relatively similar conditions). With this notion in place, clustering analysis can again be utilized on NIS-IRN to identify groups of ports based on NIS invasion risk. The clusters detected here are sub-clusters of SFN clusters (that are based on species flow dynamics); therefore, if two ports are in the same cluster of SFN, and they are in the same sub-cluster of NIS-IRN, then it is very likely for an NIS invasion to occur between this two ports. Furthermore, having frequent species exchanges and similar environmental conditions, if adequate ballast management strategies are not in place, a NIS invasion to any single port will immediately put all the other ports in the NIS-IRN sub-cluster at risk of a NIS invasion.

\section{Results and Discussion}
\label{sec:results_discuss}

\begin{figure*}[ht!]
\begin{center}
\includegraphics[width=7in]{%
	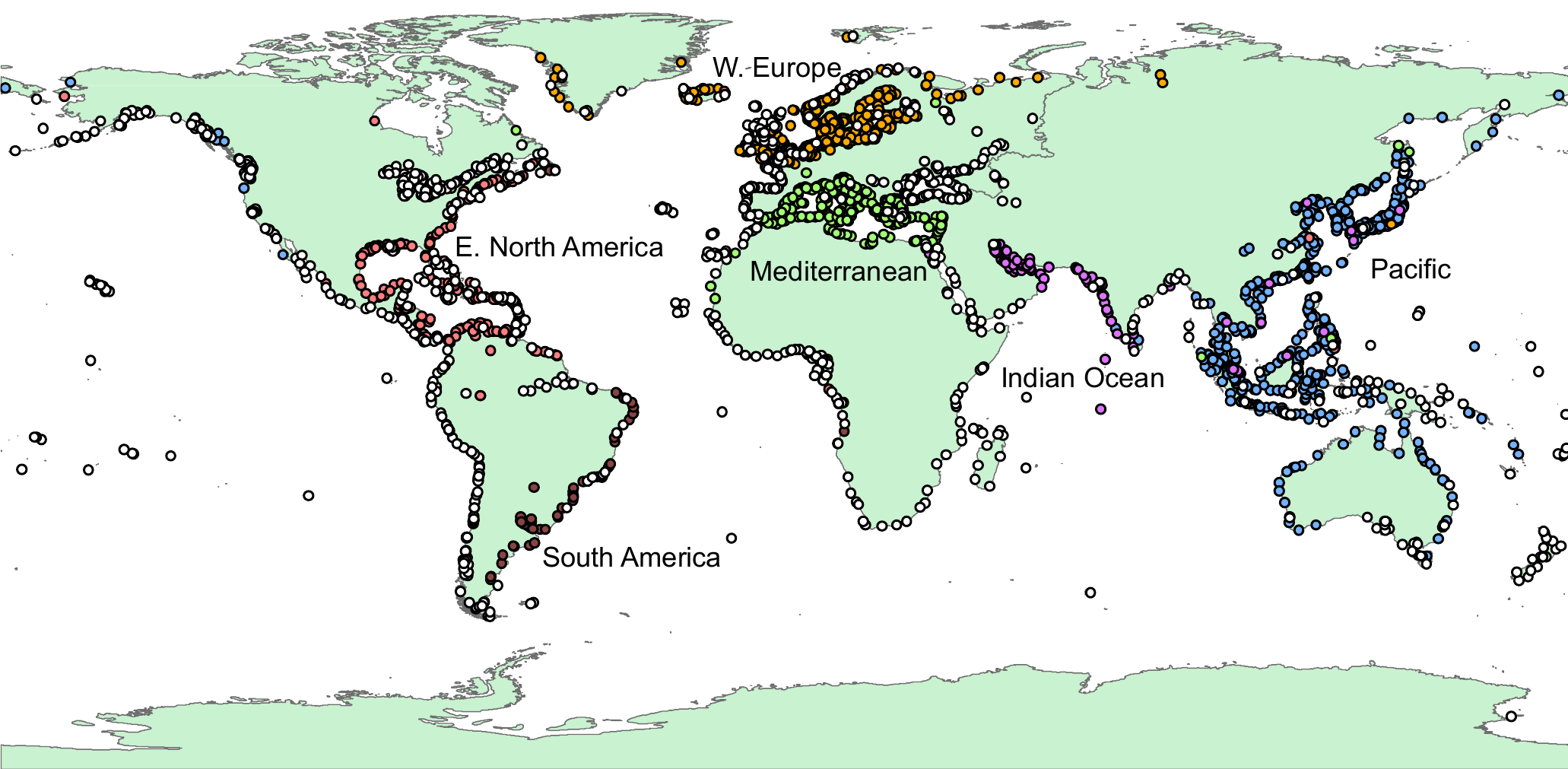}
\vspace*{-0.15in}
\end{center}
\caption{\scriptsize\tb{Major clusters of SFN during 2005--2006.} Major clusters remain largely unchanged for the duration of 1997--2006, and contain a significant proportion of total species flow between ports.}
\label{fig:2005_infomap}
\vspace*{-0.1in}
\end{figure*}

Clustering analysis of SFN reveals several clusters of ports. While clustering is derived based on species flow dynamics, geographical orientation of the major (in terms of size) clusters is also intuitive (see Fig.~\ref{fig:inter_intra_clusters}). What is more interesting is the fact that these major clusters continue to exist over the duration studied---for a given cluster, while some ports leave/join over time, the vast majority of the ports continue to remain. This perhaps provides a basis for devising efficient management strategies that are based on port clusters and inter-cluster connections. Let us now explore these observations in detail.

\subsection{Major clusters, their evolution and interactions}
A few major clusters correspond to a significant proportion of total species flow between ports. For instance, in 2005--2006, six (6) major clusters (out of 64 in total), viz., the clusters of {\tt Pacific}, {\tt Mediterranean}, {\tt Western\_Europe}, {\tt Eastern\_North\_America}, {\tt Indian\_Ocean} and {\tt South\_America} contain 68.6\% of total ports and correspond to 76.3\% of the total species flow (see Table~\ref{tb:major_cluster_ports} for ports that are of interest in each of these major clusters). 

\begin{figure}[ht!]
\begin{center}
\includegraphics[width=3.5in]{%
	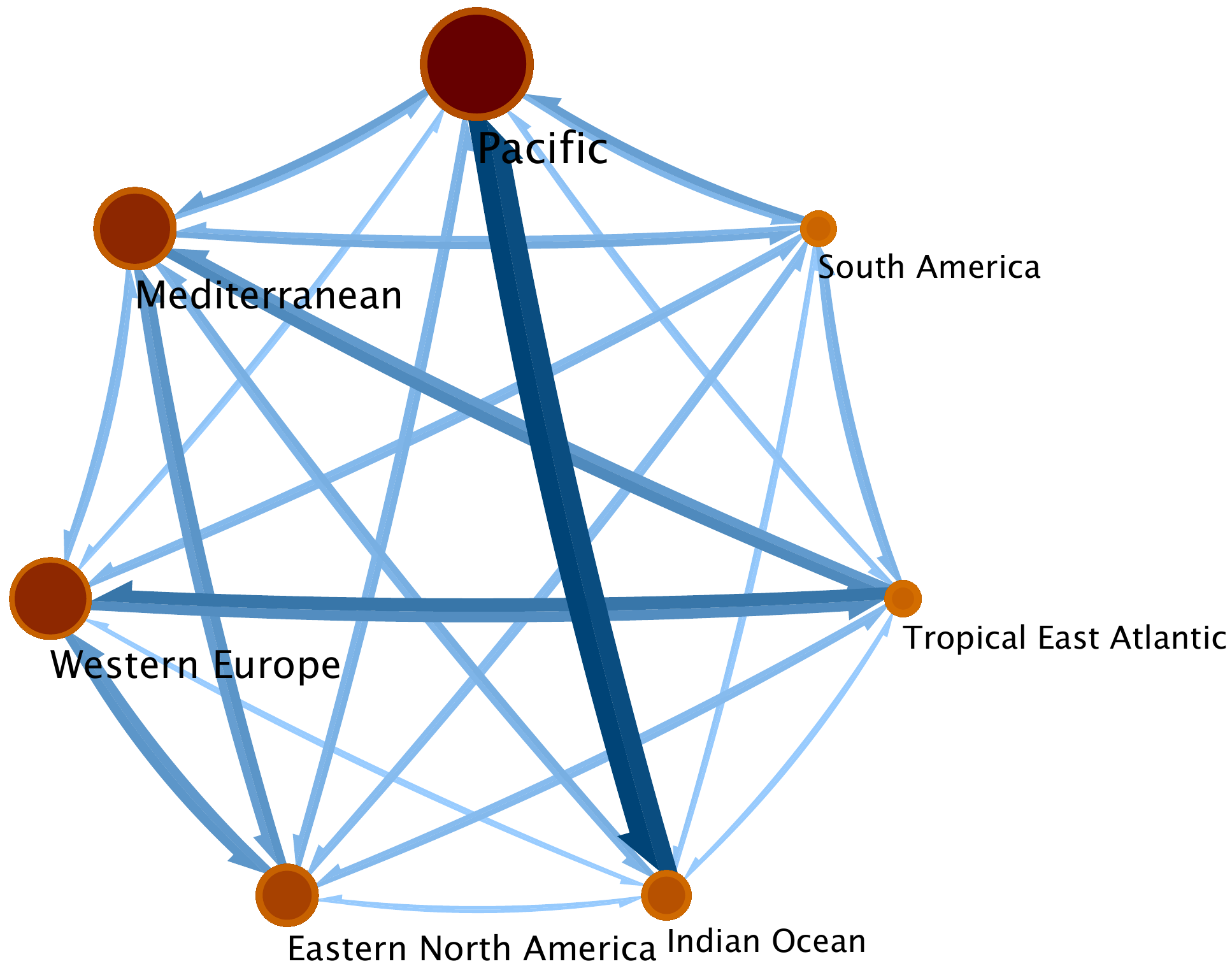}
\end{center}
\vspace*{-0.2in}
\caption{\scriptsize\tb{Seven (7) major clusters of the 2005--2006 dataset is shown with all inter-cluster connections.} Here, ratio of darker/lighter region explains the ratio of intra-cluster flow (i.e., flow between ports within a cluster) to inter-cluster flow (i.e., flow between ports belonging to different clusters). Therefore, in major clusters, species exchange among ports within clusters appears to be much higher compared to that of between clusters. Also, some clusters indicate relatively higher species exchange, i.e., between Indian Ocean and Pacific clusters.}
\label{fig:inter_intra_clusters}
\vspace*{-0.1in}
\end{figure}

\begin{table*}[ht!]
\vspace*{-0.1in}
\scriptsize
\renewcommand{\arraystretch}{1.1}\addtolength{\tabcolsep}{-4.1pt}
\caption{\tb{Ports that remain in the same cluster for the duration of 1997--2006} }
\vspace*{-0.1in}
\begin{tabular}{l|rr || l|rr || l|rr || l|rr || l|rr || l|rr}
\hline
\mcol{3}{c||}{\tb{Pacific}}
&\mcol{3}{c||}{\tb{Mediterranean}}
&\mcol{3}{c||}{\tb{W. European}}
&\mcol{3}{c||}{\tb{E. North\_America}}    
&\mcol{3}{c||}{\tb{Indian\_Ocean}}
&\mcol{3}{c}{\tb{South\_America}}\\
\mcol{3}{c||}{\%TP=28.33\%, \#P=818}
&\mcol{3}{c||}{\%TP=15.61\%, \#P=513}    
&\mcol{3}{c||}{\%TP=15.37\%, \#P=1117}
&\mcol{3}{c||}{\%TP=9.31\%, \#P=363}
&\mcol{3}{c||}{\%TP=6.12\%, \#P=137}
&\mcol{3}{c}{\%TP=3.41\%, \#P=80}\\
\hline
Port name    &\%TF    &\%CF
&Port name    &\%TF    &\%CF
&Port name    &\%TF    &\%CF
&Port name    &\%TF    &\%CF
&Port name    &\%TF    &\%CF
&Port name    &\%TF    &\%CF\\
\hline
Singapore    &2.82    &9.96
	&Gibraltar &2.56    &16.37
		&Rotterdam    &0.87    &5.68
			&Houston    &0.52    &5.57
				&Jebel Ali    &0.25    &4.07    
					&Santos    &0.42    &12.37\\
Hong Kong    &0.68    &2.41
	&Tarifa    &0.86    &5.54
		&Skaw    &0.60    &3.93
			&New Orleans    &0.37    &3.94
				&Ras Tanura    &0.22    &3.67
					&Tubarao    &0.33    &9.70\\
Kaohsiung    &0.58    &2.05
	&Port Said    &0.84    &5.38
		&Antwerp    &0.55    &3.59
			&New York    &0.35    &3.80
				&Mumbai    &0.20    &3.29
					&San Lorenzo$^*$    &0.33    &9.57\\
Port Hedland    &0.52    &1.83
	&Suez    &0.48    &3.09
		&Brunsbuttel    &0.44    &2.85
			&Baltimore    &0.23    &2.42
				&Juaymah Term.    &0.19    &3.12
					&Paranagua    &0.21    &6.11\\
Busan    &0.50    &1.76
	&Barcelona    &0.29    &1.83
		&Hamburg    &0.42    &2.76
			&Port Arthur    &0.21    &2.28
				&Kharg Is.    &0.18    &2.91
					&Rio de Janeiro    &0.15    &4.45\\
Hay point    &0.49    &1.72
	&Venice    &0.24    &1.52
		&Amsterdam    &0.31    &2.02
			&Santa Marta    &0.20    &2.17
				&Jubail    &0.17 &2.76
					&Bahia Blanca    &0.15    &4.31\\
Newcastle$^{**}$    &0.48    &1.71
	&Genoa    &0.23    &1.47
		&Immingham    &0.28    &1.83
			&Tampa    &0.20    &2.16
				&New Mangalore    &0.15    &2.50
					&Rosario    &0.14    &4.07\\
Gladstone    &0.47    &1.67
	&Piraeus    &0.22    &1.39
		&St. Petersburg    &0.27    &1.73
			&Port Everglades    &0.20    &2.13
				&Mesaieed    &0.13    &2.08
					&Sepetiba    &0.12    &3.60\\
Nagoya    &0.46    &1.61
	&Leghorn    &0.21    &1.32
		&Tees    &0.22    &1.41
			&Mobile    &0.19    &2.04
				&Bandar Abbas    &0.12    &2.03
					&Rio Grande$^{***}$    &0.12    &3.59\\
Incheon    &0.45    &1.60
	&Augusta&0.20    &1.26    
		&Zeebrugge&0.21    &1.36    
			&Savannah    &0.18    &1.95
				&Jebel Dhanna Term.    &0.12    &1.95
					&Praia Mole    &0.12    &3.50\\    
\hline
\end{tabular}
{\scriptsize Ports corresponding to highest \%TF:=percentage flow w.r.t. total flow and \%CF:=percentage flow w.r.t. flow within cluster are shown for six major clusters; for each cluster, the aggregated \%TF:=percentage flow in the cluster w.r.t. total flow and number of ports in the cluster are given in the first row of table. %
Here, San Lorenzo$^*$:=San Lorenzo, Argentina; Newcastle$^{**}$:=Newcastle, Australia;  Rio Grande$^{***}$:=Rio Grande, Brazil.}
\label{tb:major_cluster_ports}
\end{table*}

\begin{figure}[ht!]
\begin{center}
\includegraphics[width=\figwidth]{%
	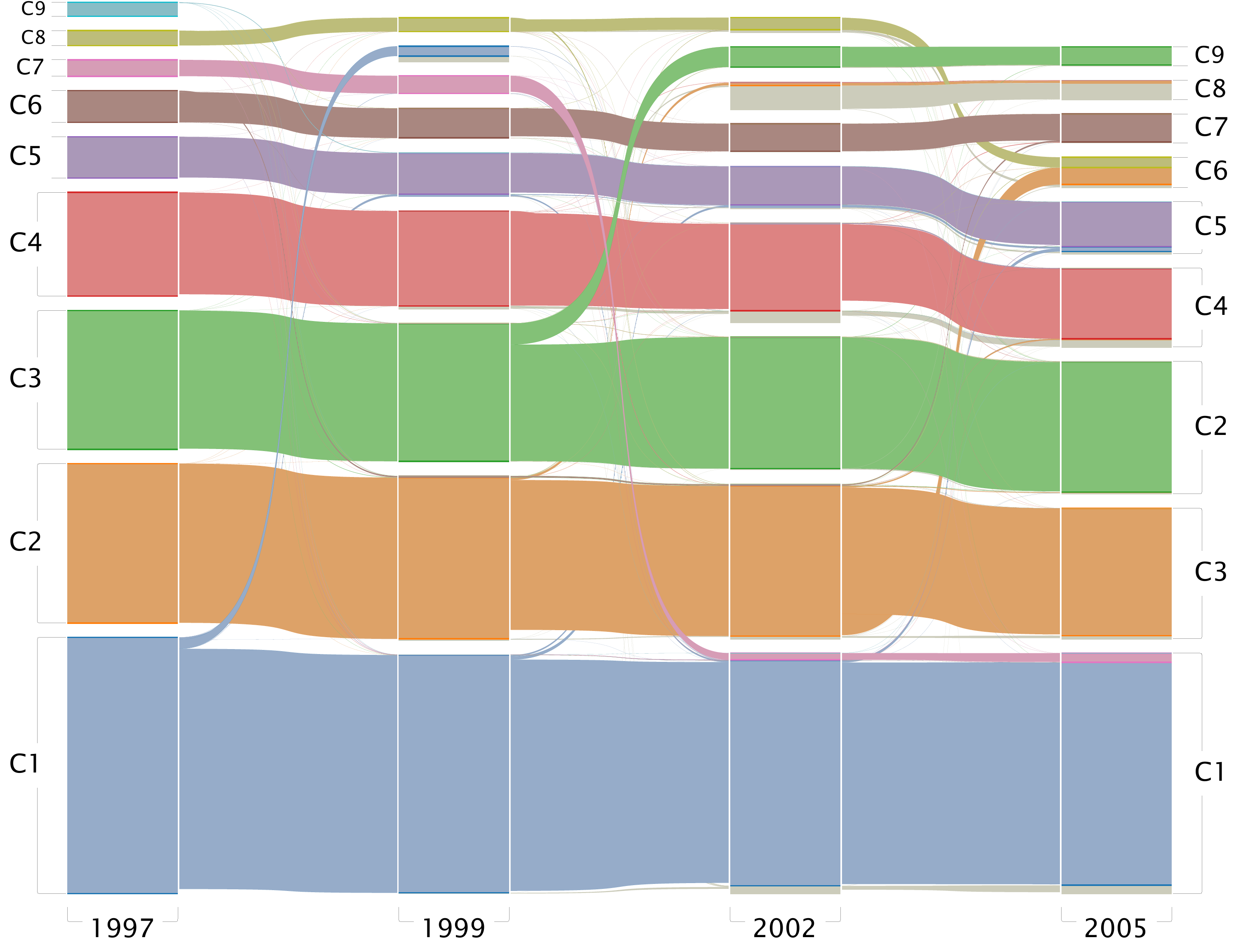}
\vspace*{-0.35in}
\end{center}
\caption{\scriptsize\tb{Illustration of evolution of major clusters during the period of 1997--2006.} The clusters in \emph{alluvial diagram}~\cite{Ros08} are ranked by aggregated flow within the cluster. Here, the columns 1997, 1999, 2002 and 2005 represent the major clusters of SFN generated for LMIU datasets for 1997--1998, 1999--2000, 2002--2003 and 2005-2006, respectively. With respect to 2005--2006 clusters, C1:= {\tt Pacific}, C2:= {\tt Mediterranean}, C3:= {\tt Western\_Europe}, C4:= {\tt Eastern\_North\_America}, C5:= {\tt Indian\_Ocean}, C6:= {\tt Tropical\_East\_Atlantic}, and C7 = {\tt South\_America}. Note that the ranking of {\tt Mediterranean} and {\tt Western\_Europe} has exchanged from 2002--2003 to 2005--2006.}
\label{fig:alluvial}
\vspace*{-0.1in}
\end{figure}

The evolution of clusters (i.e., how ports get grouped over time) can reveal important information on how changes in vessel movement (and ballast discharge) patterns affect species flow dynamics (see Fig.~\ref{fig:alluvial}). Perhaps, one of the most trivial transitions being observed is the exchange of the order of the two clusters {\tt Mediterranean} (contains the ports of Baltic Sea, Celtic Seas, North Sea, Norway, SW Iceland, West Greenland Shelf, etc.) and {\tt Western\_Europe} (contains the ports of Alboran Sea, Adriatic Sea, Aegean Sea, Ionian Sea, West Mediterranean, Levantine Sea, etc.) from 2002--2003 to 2005--2006. Here, the clusters in \emph{alluvial diagram}~\cite{Ros08} are ranked by aggregated flow within the cluster; therefore, what's being observed is a relative increase of species exchange among ports that belong to these clusters during 2005--2006. This change can perhaps be attributed to the merger of a significant proportion of ports belonging to {\tt Mediterranean} cluster (including Celtic Seas and South European Atlantic Shelf) with C9 (includes Azores Canaries Madeira, Saharan Upwelling, South European Atlantic Shelf, etc.) in 2002--2003 to form the {\tt Tropical\_East\_Atlantic} cluster in 2005--2006. 

Several ports that isolate themselves from the major clusters to form a smaller cluster (or clusters) are also observed. For instance, 21 ports in California and Hawaii, including ports such as San Francisco, Los Angeles and San Diego that belonged to the {\tt Pacific} cluster in 1997--1998 form a new smaller cluster (C8 in Fig.~\ref{fig:alluvial}) in 1999--2000. Similarly, port of Miami that belonged to {\tt Eastern\_North\_America} from 1997 to 2003 joins a small cluster (C16) in 2005--2006 indicating that the port of Miami and some other ports in Bahamian, Floridian, Greater Antilles and Western Caribbean no longer belong to the cluster of {\tt Estern\_North\_America} (contains ports such as Carolinian, Floridian, Greater Antilles, Guinan, etc.). On the contrary, some ports that previously belonged to smaller clusters, merge with major clusters. This indicates a significant increase in vessel movement between these ports and the ports in the merged cluster. For example, the merger of {\tt IndoPacific} cluster (i.e., C7 in alluvial diagram under 1997--2000) with {\tt pacific} in 2002--2003 is a clear example, where the ports Java Sea and Malacca Strait are part of this movement. 

\subsection{Species flow control and efficient ballast management}
\begin{figure}[ht!]
\begin{center}
\includegraphics[width=\figwidth]{%
	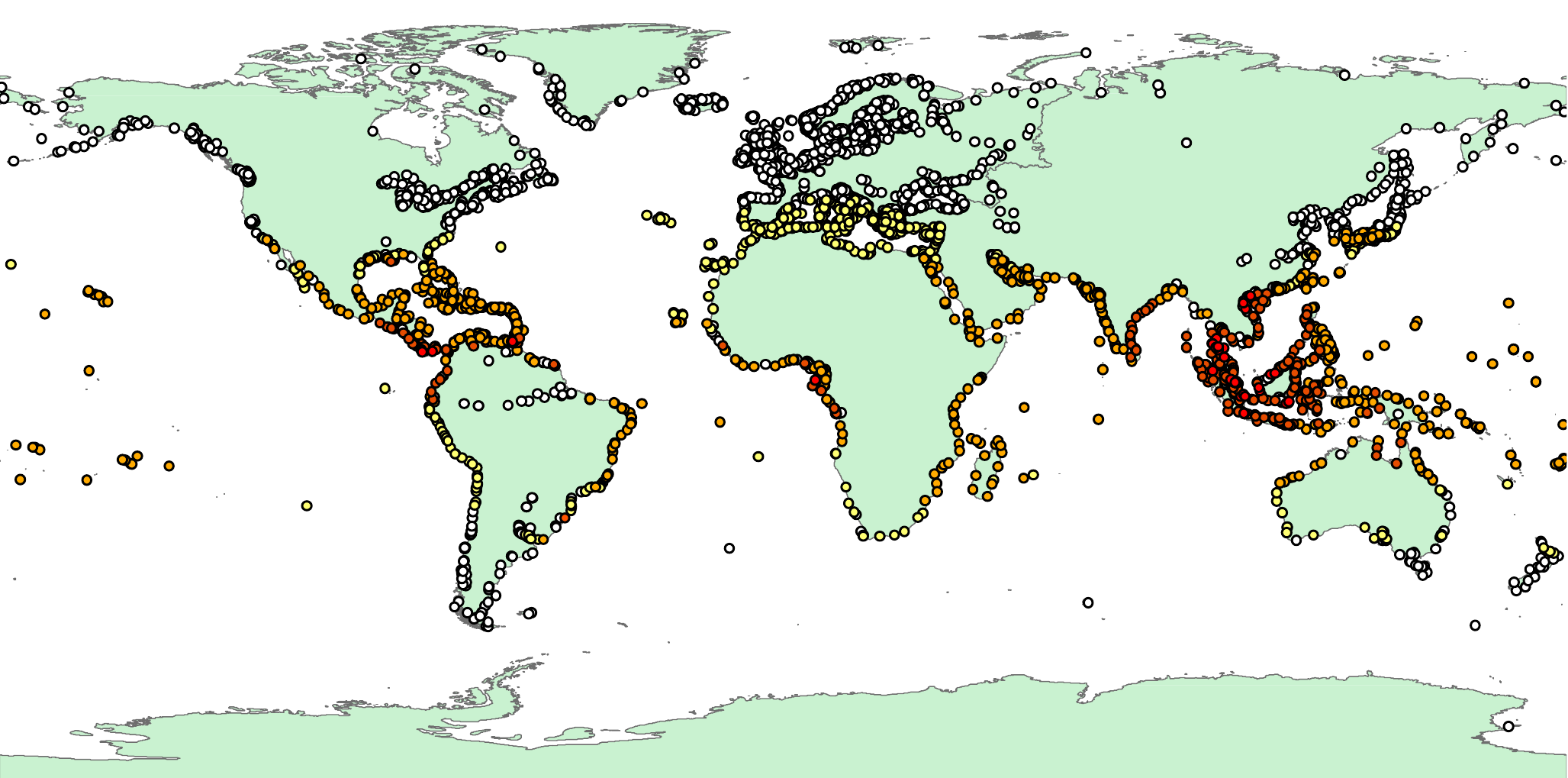}
\vspace*{-0.35in}
\end{center}
\caption{\scriptsize\tb{NIS invasion risk with respect to Singapore.} The colors correspond to risk level definition introduced in Fig.~\ref{fig:risk_level} (b); here, white denotes risk level zero (0), or no invasion risk.} 
\label{fig:Singapore_RiskLevel}
\vspace*{-0.1in}
\end{figure}

Species exchange among ports can be efficiently controlled by targeted management on pathways that isolate ports (or clusters of ports) from others. Let us consider Fig.~\ref{fig:inter_intra_clusters} that illustrates the inter- and intra-cluster species flow among major clusters. Clearly, most of the clusters have relatively higher species exchange among ports within the cluster compared to ports that do not belong the same cluster. Furthermore, while inter-cluster connections change, this pattern is virtually consistent over time. Therefore, species introduction pathways can be combinatorially reduced via targeted management on inter-cluster connections. %
\begin{figure}[ht!]
\begin{center}
\subfigure[1997-1998]{
\includegraphics[width=1.5in]{%
	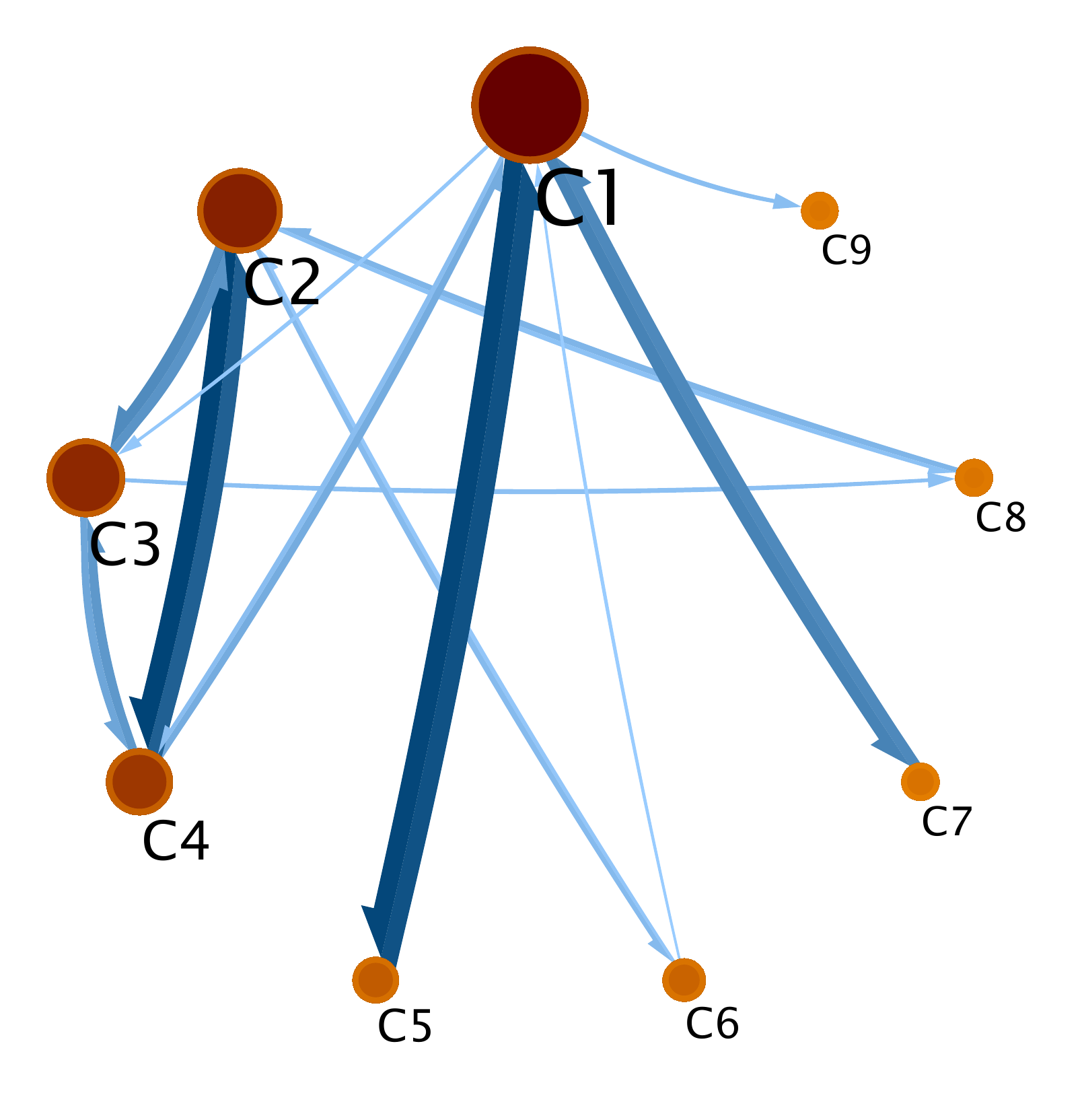}}\hfill
\subfigure[1999-2000]{
\includegraphics[width=1.5in]{%
	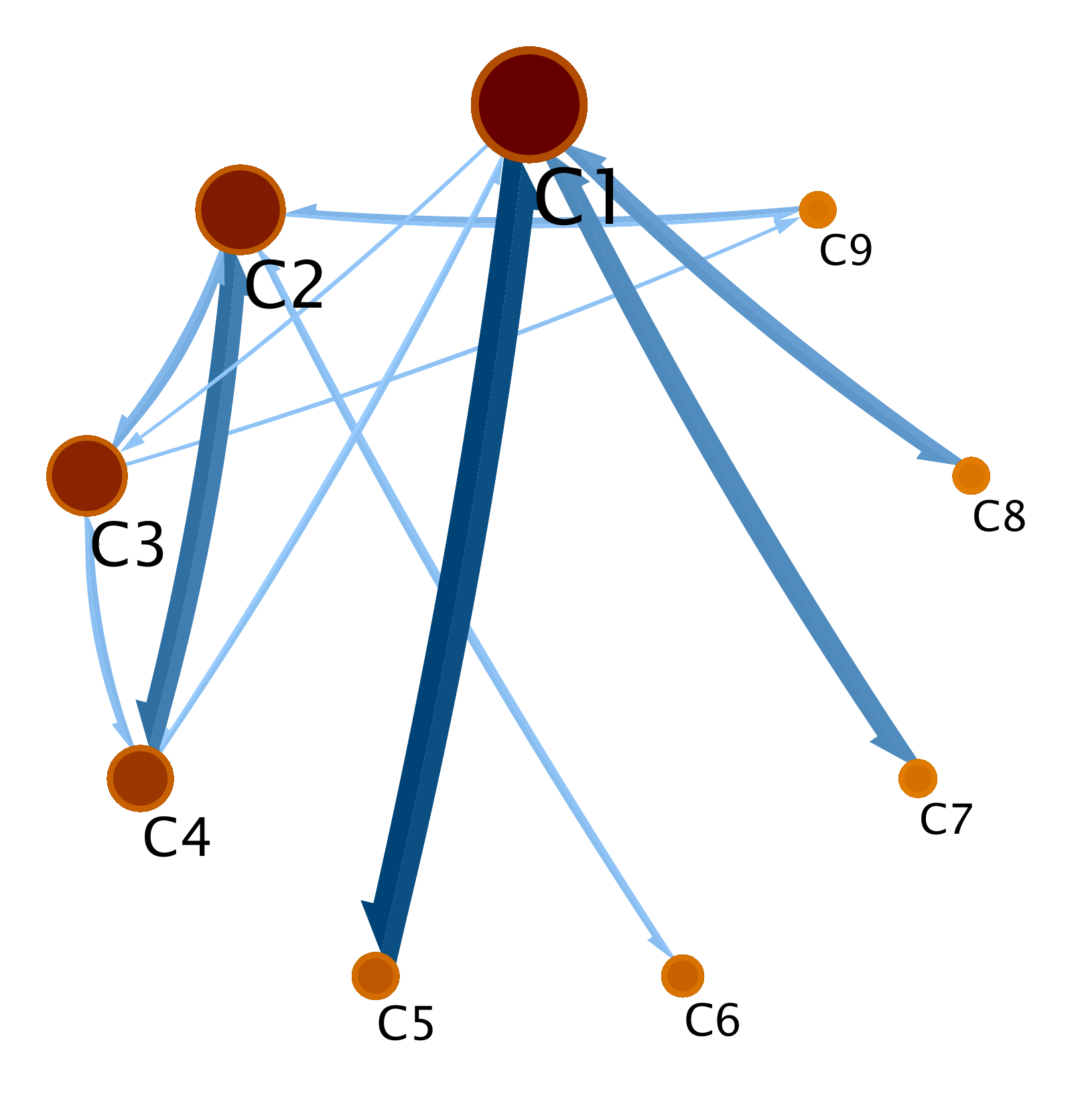}}
\subfigure[2002-2003]{
\includegraphics[width=1.5in]{%
	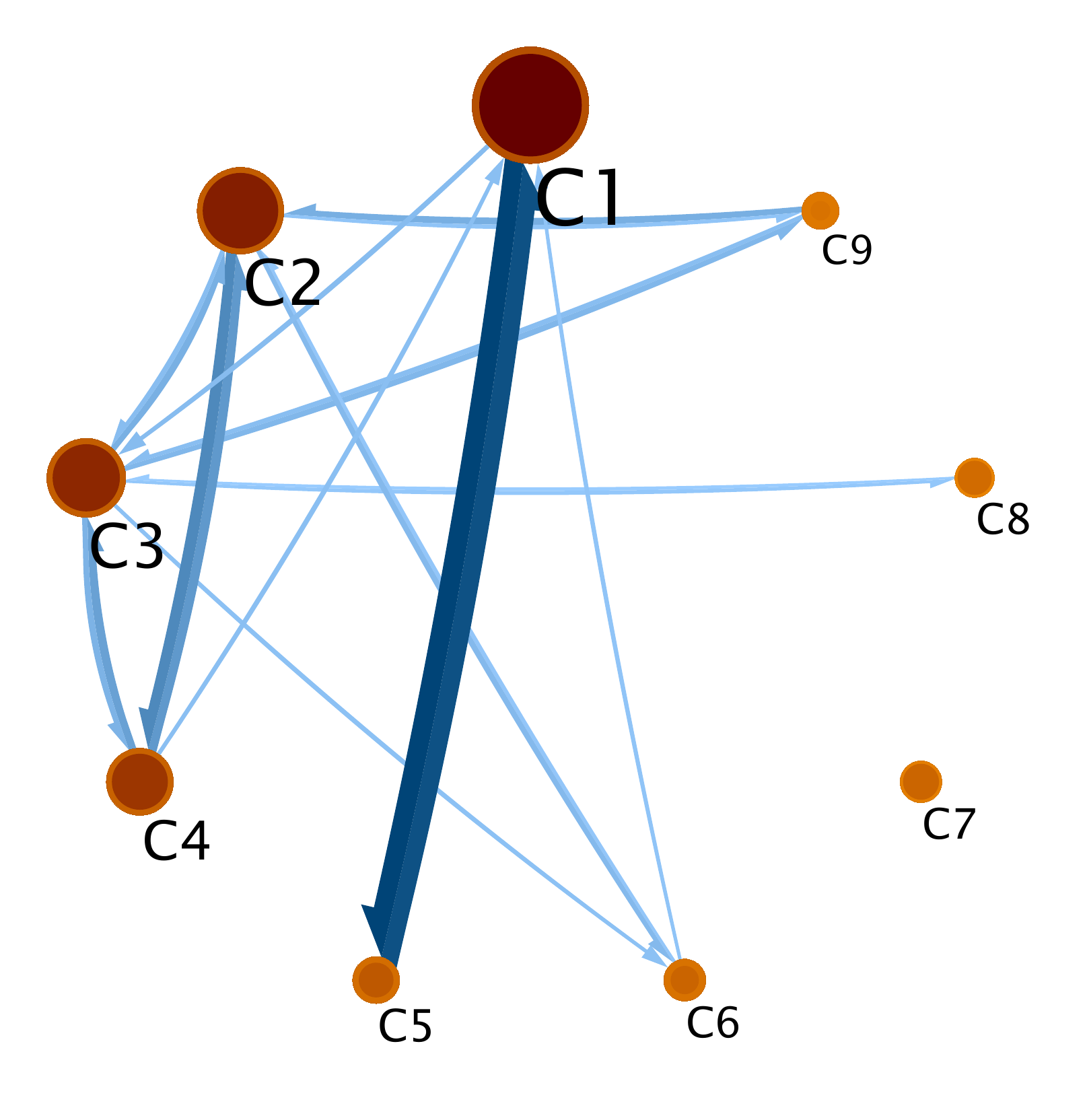}}\hfill
\subfigure[2005-2006]{
\includegraphics[width=1.5in]{%
	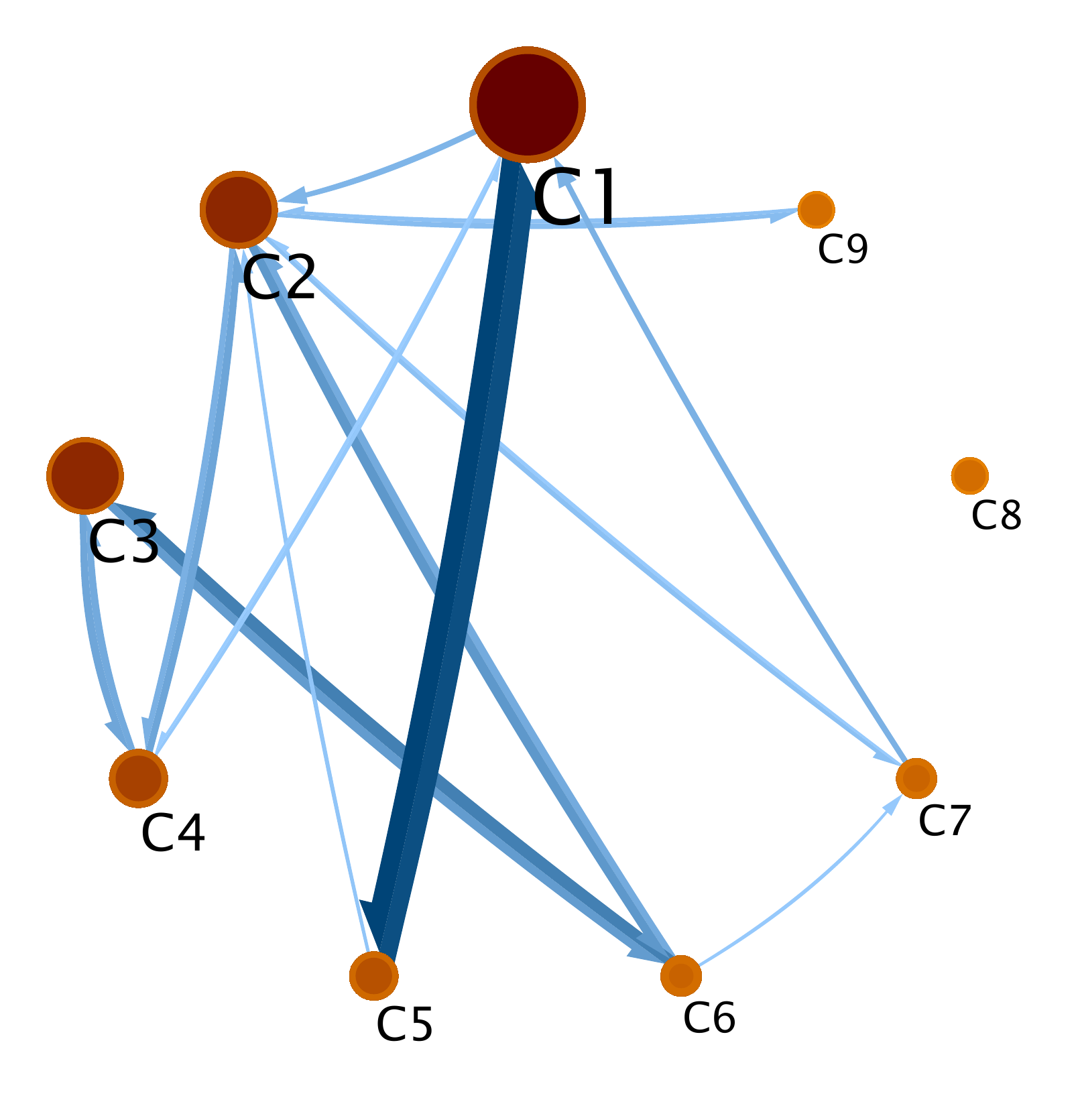}}
\end{center}
\vspace*{-0.1in}
\caption{\scriptsize\tb{Illustration of inter-cluster and intra-cluster flow.} Here, ratio of darker/lighter region explains the ratio of intra-cluster flow (i.e., flow between ports within a cluster) to inter-cluster flow (i.e., flow between ports belonging to different clusters). Therefore, in major clusters, species exchange among ports within clusters appears to be much higher compared to that of between clusters.}
\label{fig:inter_intra_clusters}
\vspace*{-0.1in}
\end{figure}
For instance, consider the {\tt Pacific} cluster (C1) in year 2005--2006. There are 37,596 inter-cluster connections, where Table~\ref{tb:inter_cluster_1} tabulates the strongest connections. %
\begin{table}[ht!]
\renewcommand{\arraystretch}{1.1}\addtolength{\tabcolsep}{0pt}
\parbox{.5\linewidth}{
\caption{\tb{Inter-cluster flow for Pacific cluster (C1) in 2005--2006}}
\begin{tabular}{l|l}
\hline
\tb{From Port}	& \tb{To Port}\\
\hline
Singapore 		&Port Said\\
Singapore 		&Richards Bay\\
Mormugao 		&Singapore\\
Suez     			&Singapore\\
Singapore 		&Visakhapatnam\\
Paradip 			&Singapore\\
Visakhapatnam 	&Singapore\\
Tubarao 		&Singapore\\
Chennai 		&Singapore\\
Ponta da Madeira 	&Singapore\\
\hline
\end{tabular}
\label{tb:inter_cluster_1}}
\hfill
\parbox{.45\linewidth}{
\caption{\tb{Major inter-cluster contributors for species flow in 2005--2006}}
\begin{tabular}{c|l}
\hline
\tb{Cluster}	&\tb{Port}\\
\hline
C1 		&Singapore\\
C2 		&Gibraltar\\
C3 		&Rotterdam\\
C4  		&New York\\
C5  		&Mormugao\\
C6  		&Cape Finisterre \\
C7  		&Tubarao\\
C8  		&Seven Islands\\
C9  		&Istanbul\\
C10  	&Long Beach\\
\hline
\end{tabular}
\label{tb:inter_cluster}}
\end{table}
Clearly, Singapore is in the focus; in fact, Singapore alone contributes to approximately 26\% of total inter-cluster flow from/to {\tt Pacific} cluster that contains 818 ports (see Fig.~\ref{fig:Singapore_RiskLevel} for an illustration of NIS invasion risk with respect to Singapore). Here, via targeted ballast management on inter-cluster connections to/from Singapore and few other ``influential'' ports, inter-cluster flow from/to {\tt Pacific} cluster can be significantly reduced with a minimal effort. 

Consider Table~\ref{tb:inter_cluster} that lists ports corresponding to the highest inter-cluster flow in ten (10) major clusters for 2005--2006. This table gives us an effective approach for inter-cluster ballast water control for ports: by advocating mandatory ballast water control on these few ports, a large proportion of inter-cluster connections can be eradicated, thus greatly reducing the risk of inter-cluster NIS invasion.

%

\subsubsection{Impact of ballast management on highly connected ports}
Average path length of three (3) that is observed on SFN indicates that species could be translocated between any two given ports within two (2) stopovers on average. This indicates an extremely viable situation for NIS exchange, especially if proper species management strategies are not in place. In order to understand the impact of targeted ballast management on average path length, a test scenario based on a hypothetical SFN---$\widehat{\tx{SFN}}$ can be derived as follows:
	\tb{(i)} choose an SFN (SFN corresponding to 2005--2006 LMIU dataset was chosen for our experiment); then,
	\tb{(ii)} identify 20\% of all ports with the highest degree (see Table~\ref{tab:highest_degree_ports}); and, finally
	\tb{(iii)} generate $\widehat{\tx{SFN}}$ by removing all edges to/from the above ports; this corresponds to ballast management with 100\% efficiency, i.e., zero (0) species flow from/to these ports. 
\emph{Then, the average path length increases to 6.4 indicating that it will be twice as difficult for species to be translocated from one port to another.} 
Furthermore, higher average path length also implies, 
	\tb{(i)} longer travel times (hence, very low chance of survival for species during the voyage) and 
	\tb{(ii)} increased number of intermediate stop-overs (which is likely to dilute ballast water and expose to multiple treatments). 

\begin{table}[ht!]
\renewcommand{\arraystretch}{1.1}\addtolength{\tabcolsep}{2pt}
\caption{\tb{Ports$^*$ with degree $> 1000$ in 2005-2006 SFN}}
\vspace*{-0.1in}
\begin{center}
\begin{tabular}{l|c|l}
\hline
Port name 			& Degree 	&\hfil Important pathways (connected ports)\\
\hline
Gibraltar				& 1882	&Cape Finisterre, Tubarao\\
Dover Strait$^*$		& 1747 	&Cape Finisterre, Rotterdam, Tubarao\\
Singapore				& 1569	&Mormugao, Tubarao\\
Cape Finisterre			& 1387	&Gibraltar, Rotterdam, Tubarao\\
Panama Canal$^*$		& 1275	&New Orleans\\
Tarifa 				& 1224	&Gibraltar, Cape Finisterre\\
Rotterdam				&1126	&Cape Finisterre, Dover Strait\\
\hline
\end{tabular}
\end{center}
\vspace*{0.01in}
{\scriptsize $^*$ indicates locations in LMIU database, but do not correspond to actual ports; connected ports are listed in decreasing order of degree.}
\label{tab:highest_degree_ports}
\end{table}

\subsubsection{Vessel types and species flow}
The exact amount of species relocated by a vessel depends on many factors: ballast size, average duration per trip, frequently visited ports, etc. Furthermore, vessel types we observe in GSN are often chosen for specific tasks (e.g., oil transportation, vehicle transportation, etc.) and these vessels often have frequent and favorite ports/routes. Therefore, we investigate the relationship of vessel types to inter- and intra-cluster species flow in order to understand any existing patterns that maybe helpful in devising species management strategies (based on latest 2005-2006 LMIU dataset). 

\paragraph{Frequent inter-cluster travelers}
While not being the most frequent, {\tt container carriers} correspond to 57,909, or equivalently 24\% of all inter-cluster trips in 2005-2006. Among the most frequent vessel types, {\tt bulkers}, {\tt crude oil tankers}, {\tt refrigerated general cargo ships} and {\tt combined bulk and oil carriers} tend to travel inter-cluster for over 25\% of the time. Furthermore, among the vessel types that do not travel frequently, some vessel types tend to travel inter-cluster in majority of their trips (e.g., {\tt wood-ship carriers}: 40.4\%, {\tt livestock carriers}: 34.3\%, {\tt semi-sub HL vessels}: 37.4\% and {\tt barge container carriers}: 55.7\%).  

\paragraph{Frequent intra-cluster travelers}
Among the most frequent vessel types, {\tt passenger carriers} tend to stay within clusters for 97.6\% of their trips, thus imposing only a very minimal risk in terms of inter-cluster species translocation. Similarly, {\tt barge ships} also stay within the cluster for 98.1\% of total trips. 

\subsection{Impact of environmental conditions on invasion risk}
Cluster analysis on SFN simplifies the GSN to port clusters and inter-cluster connections. By enforcing strict ballast management on major inter-cluster connections, species exchange from one cluster to another can be very efficiently reduced. However, management of species exchange within a cluster is more difficult due to relatively high traffic between ports (and hence expensive). However, as it turns out, majority of the ports that belong to a given cluster belong to same or adjacent ecoregions. Therefore, irrespective of the magnitude, such species exchanges do not cause NIS invasions (see Section~\ref{sssec:nis_invasion_risk}). Furthermore, species exchanges that occur among ports with sufficiently high difference in environmental conditions also do not cause NIS invasions. The presented clustering analysis on NIS-IRN exploits these facts to identify sub-clusters of ports within a (SFN) cluster, in order to identify mandatory pathways that needed to be managed for efficient NIS flow control.

\begin{figure*}[htb!]
\begin{center}
\subfigure[Subclusters based on NIS-IRN]{
\includegraphics[width=2in]{%
	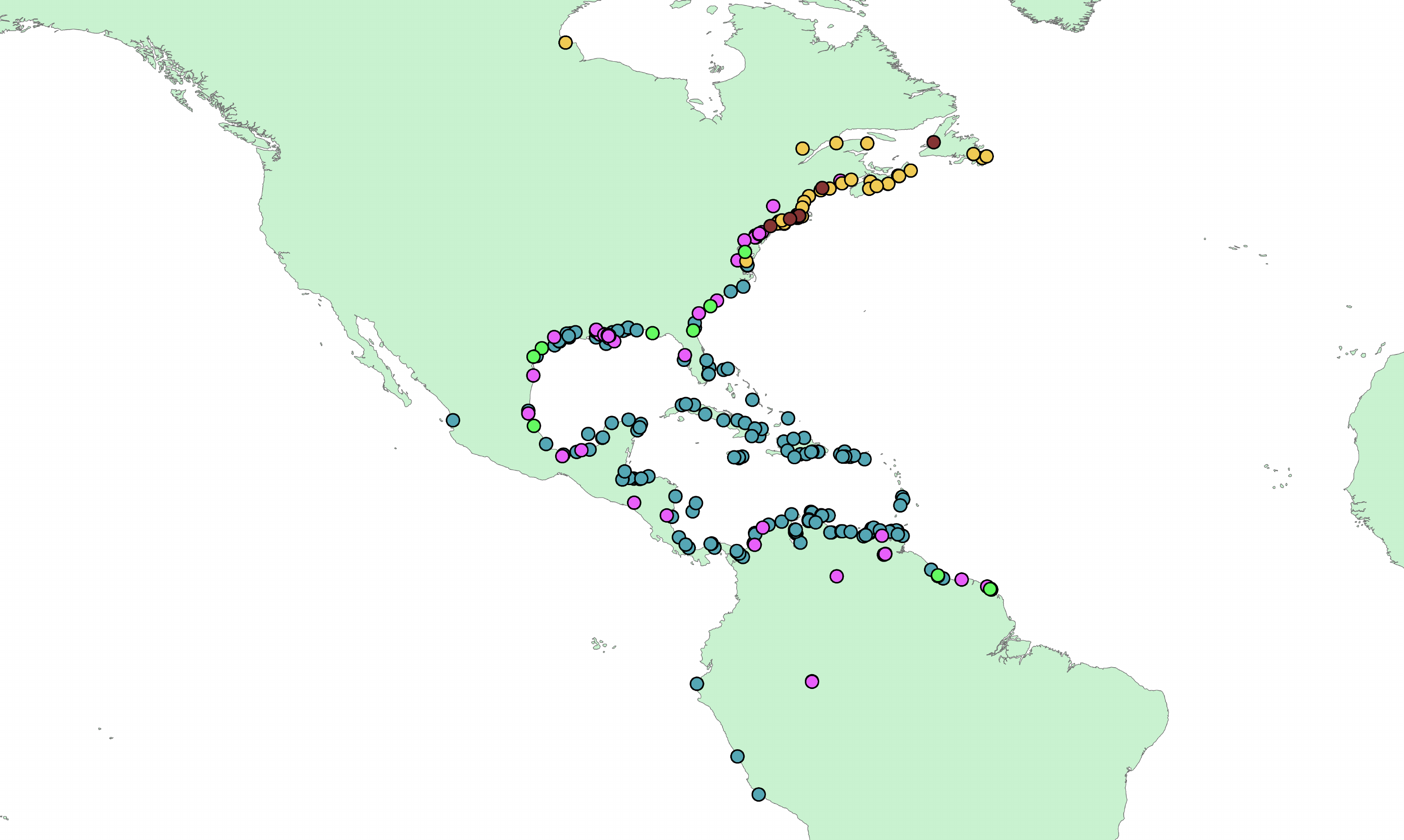}
\label{fig:env_subcluster1}	
}\hfill
\subfigure[Average Temperature]{
\includegraphics[width=2in]{%
	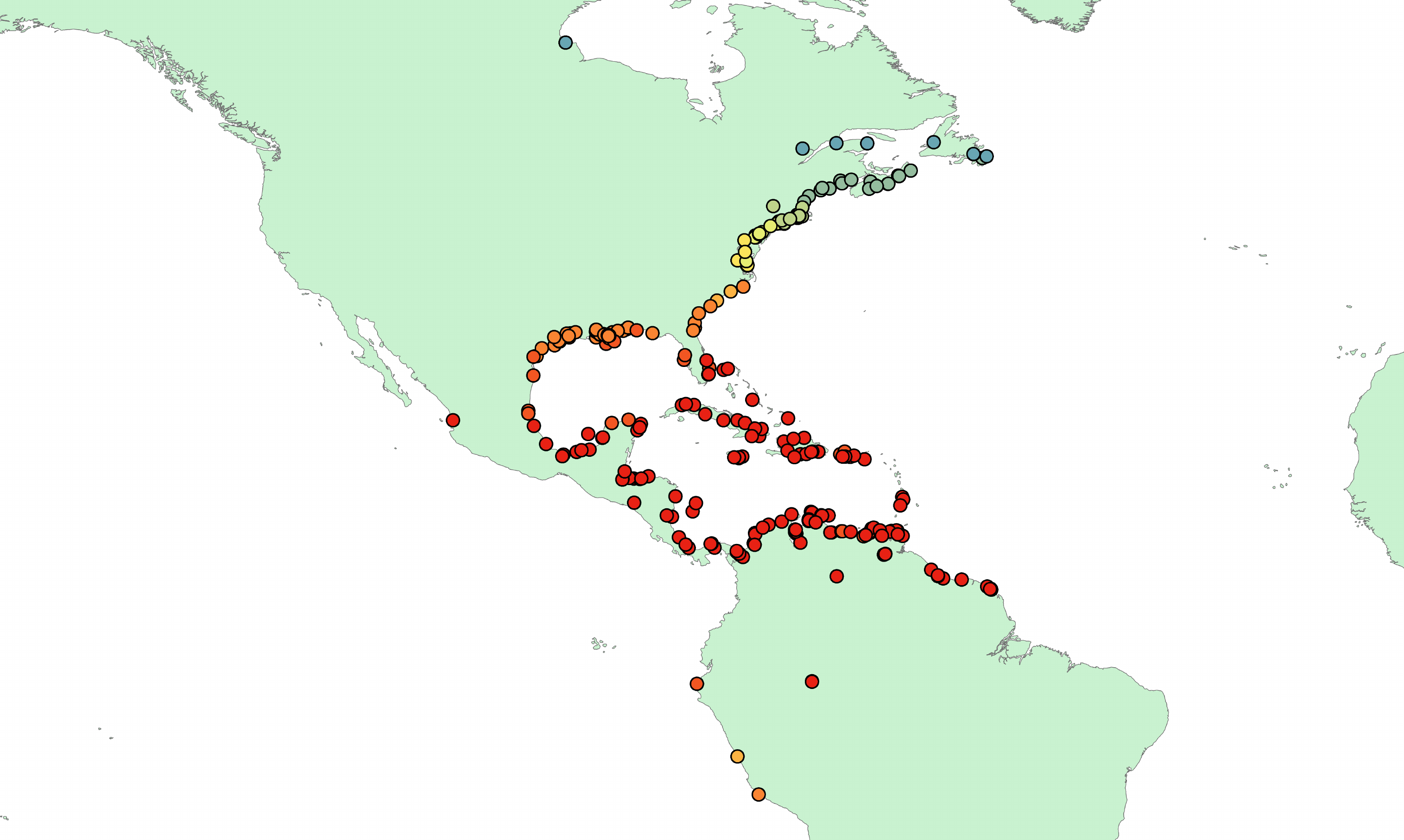}
\label{fig:env_subcluster2}	
}\hfill
\subfigure[Average Salinity]{
\includegraphics[width=2in]{%
	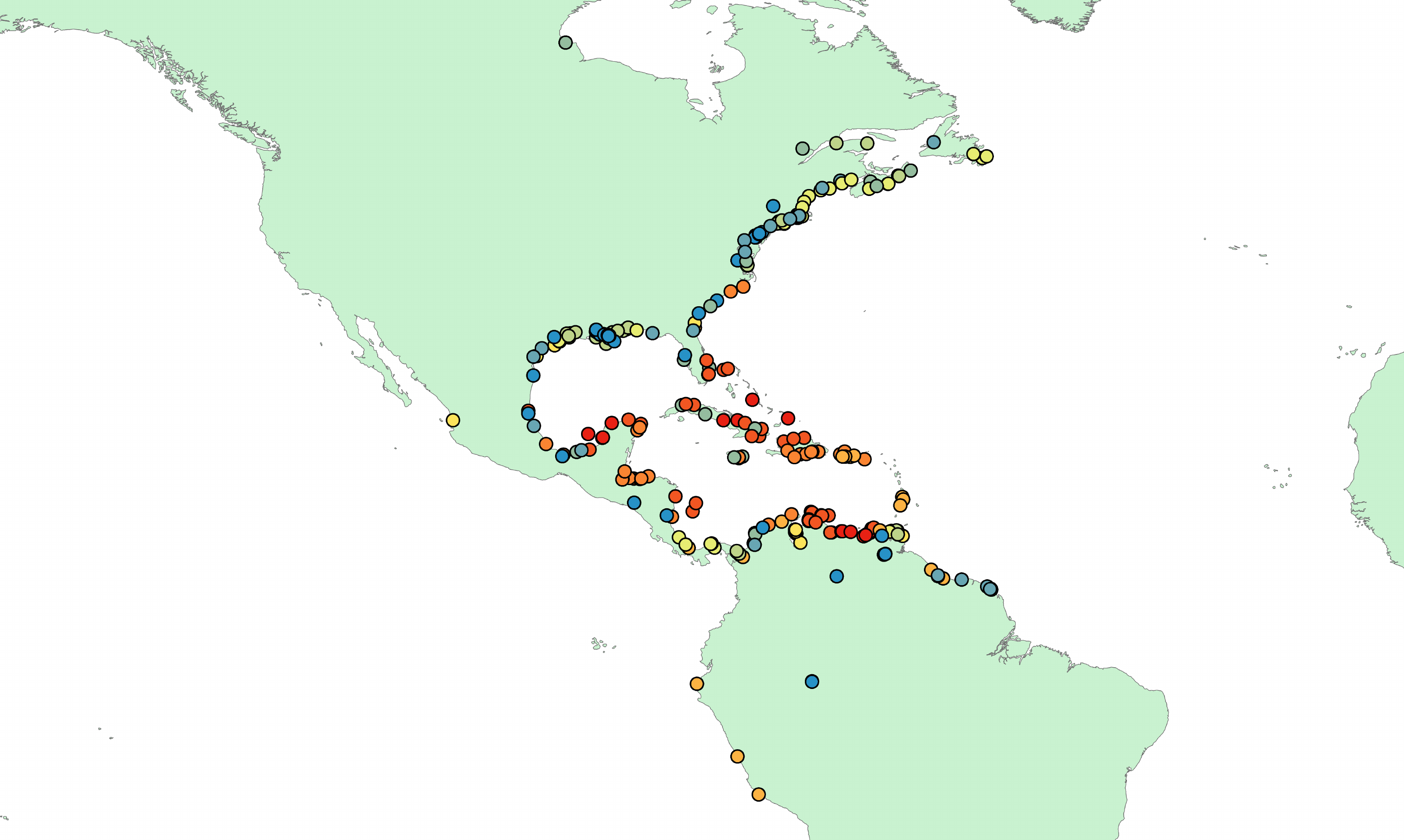}
\label{fig:env_subcluster3}	
}\hfill
\end{center}
\vspace*{-0.1in}
\caption{\tb{Result of environmental similarity network analysis in {\tt Eastern\_North\_America} cluster in 2005--2006.} }
\label{fig:env_subcluster}
\vspace*{-0.1in}
\end{figure*}

To illustrate, let us consider {\tt Eastern\_North\_America} cluster, where majority of the ports are located in north America. Five (5) sub-clusters are identified on NIS-IRN of {\tt Eastern\_North\_America} cluster; see Fig.~\ref{fig:env_subcluster1}. Average Temperature and Salinity of ports in {\tt Eastern\_North\_America} cluster are shown in Fig.~\ref{fig:env_subcluster2} and Fig.~\ref{fig:env_subcluster3}, respectively. This approach effectively identifies groups of ports such that the ports within a sub-cluster have similar environmental conditions, but are very different to the ports that do not belong to it; this can be clearly seen by comparing sub-clusters to temperature and salinity maps. In particular, ports are clustered into 5 groups: 
	(i) ports that are marked in ``blue'' are approximately $26.0\,^{\circ}\mathrm{C}$, $33.6\,\mathrm{ppt}$ which are typical of marine ports near the equator;
	(ii) ports that are marked in ``pink'' are approximately $21.6\,^{\circ}\mathrm{C}$, $1.7\,\mathrm{ppt}$ which have significantly lower salinity, that they are either fresh water ports or estuaries.
	(iii) ports that are marked in ``yellow'' are approximately $8.7\,^{\circ}\mathrm{C}$, $29.9\,\mathrm{ppt}$; most of which are to the northeast of Virginia, with a significantly lower average temperature;
	(iv) ports that are marked in ``green'' are approximately $23.0\,^{\circ}\mathrm{C}$, $15.3\,\mathrm{ppt}$, where they have warm weather and moderate salinity; and
	(v) ports that are marked in ``brown'' have around $10.8\,^{\circ}\mathrm{C}$, $13.3\,\mathrm{ppt}$, where most of them lie around Virginia, where they have cold water temperature and moderate salinity. 

With significantly different environmental conditions in different sub-clusters (e.g., those ones in fresh water and sea water), it is less likely that NIS invasions will occur across these sub-clusters. On the other hand, for ports in the same sub-cluster, it is very likely that invasions will happen among those ports, since they have very similar environmental conditions, and they are tightly coupled by ballast water exchange. In this way, clustering in NIS-IRN can serve as a powerful tool to help us better predict intra-cluster invasions.

\section{Conclusion}
\label{sec:conc}
The clustering approach we present here allows us to see large-scale patterns of NIS spread through the global shipping network. We find that, with a few exceptions, major clusters and inter-cluster connectors are consistent over many years, suggesting this information could be useful for informing policies. Since our findings are qualitative, we cannot provide specific risk probabilities to any particular port or shipping route nor can we statistically evaluate our results. However, given the poor quality of invasion data, particularly date of invasion and source region, rigorous statistical tests of any invasion predictions would be difficult. For example, many invasive species are not detected for years or decades after they are introduced, due to either the difficult taxonomic nature, population growth time lags (Sakai et al. 2001) or biased taxonomic or geographic sampling effort (Ruiz et al. 2000). Further, many invasions often cannot be attributed to a single vector or source of origin (Ruiz et al. 2000, Hewitt et al. 2004, Zenetos et al. 2013).  

We do, however, find that our results correspond well with reliable invasion data available from well-sampled ports and regions. For example, the recent influx of north Pacific species into southeast Australia's Port Phillip Bay (including a goby, green algae, and several crustaceans; Hewitt et al. 2004, Lockett and Gormon 2001) aligns with our finding that this port was in large Pacific cluster from at least 1997-2006. Further, several species of Northwest African origin have been recently detected in the Mediterranean and Northeast Atlantic (Zenetos et al. 2013, Clemente et al. 2013), corresponding with increasing number of North African ports joing these clusters over time. Finally, the cluster analysis predicts that secondary spread of NIS will be high among ports within a cluster, a phenomenon that has been observed several times in both the well-studied Mediterranean and West Atlantic clusters (Leppakoski and Olenin 2000, Zenetos et al. 2013). 

Overall, our patterns corresponds with observed data well, suggesting our results could be useful for predicting NIS spread in poorly-studied regions or in the near-future. In particular, we observed an increasing number of ports in the northwest Indian Ocean have joined the Pacific cluster, highlighting the high likelihood of recent invasions between these two regions. We also note a decline in highly-connected ports along the western United States (ie. many California ports are not in a major cluster by 2005). This pattern, in tandem with tightening ballast water restrictions by California, suggests that ship-born invasion rates should be decreasing in this region. Finally, we note that Chinese ports are diversifying into non-Pacific clusters (3 non-Pacific Cluster ports in 1997 versus 9 in 2005), suggesting that this country could be a new globally important source of invasive species.


\section*{Acknowledgement}
This work is based on research supported by the ND Office of Research via Environmental Change Initiative (ECI).

\end{document}